\documentclass[fleqn,usenatbib]{mnras}

\usepackage{newtxtext,newtxmath}
\usepackage{multirow}
\usepackage{booktabs}
\usepackage{ulem}

\usepackage[T1]{fontenc}

\DeclareRobustCommand{\VAN}[3]{#2}
\let\VANthebibliography\thebibliography
\def\thebibliography{\DeclareRobustCommand{\VAN}[3]{##3}\VANthebibliography}

\usepackage{graphicx}	
\usepackage{subcaption}
\usepackage{amsmath}	
\usepackage{amssymb}	
\usepackage{cellspace}
\usepackage{makecell}
\title[LP cosmological application]{Cosmological constraints with the linear point from the BOSS survey}

\author[M. He et al.]{
Mengfan He,$^{1,2,3}$\thanks{E-mail: mfhe@bao.ac.cn}
Cheng Zhao,$^{4}$\thanks{E-mail: cheng.zhao@epfl.ch}
and Huanyuan Shan$^{2,3}$\thanks{E-mail: hyshan@shao.ac.cn}
\\
\small 
$^{1}$National Astronomical Observatories, Chinese Academy of Sciences, Beijing 100012, China\\
\small $^{2}$Shanghai Astronomical Observatory, Chinese Academy of Sciences, 80 Nandan Road, Shanghai 200030, China\\
\small $^{3}$School of Astronomy and Space Science, University of Chinese Academy of Sciences, Beijing 100049, P.R. China\\
\small $^{4}$Institute of Physics, Laboratory of Astrophysics, \'Ecole Polytechnique F\'ed\'erale de Lausanne (EPFL), Observatoire de Sauverny, CH-1290 Versoix, Switzerland\vspace*{-2pt} \\
}

\date{Accepted XXX. Received YYY; in original form ZZZ}

\pubyear{2015}

\begin{document}
\label{firstpage}
\pagerange{\pageref{firstpage}--\pageref{lastpage}}
\maketitle

\begin{abstract}
{The {\it Linear Point} (LP), defined as the midpoint between the BAO peak and the associated left dip of the two-point correlation function (2PCF), $\xi(s)$, is proposed as a new standard ruler which is insensitive to nonlinear effects. 
In this paper, we use a Bayesian sampler to measure the LP and estimate the corresponding statistical uncertainty, and then perform cosmological parameter constraints with LP measurements. 
Using the Patchy mock catalogues, we find that the measured LPs are consistent with theoretical predictions at 0.6 per cent level. 
We find constraints with midpoints identified from the rescaled 2PCF ($s^2 \xi$) more robust than those from the traditional LP based on $\xi$, as the BAO peak is not always prominent when scanning the cosmological parameter space, with the cost of 2--4 per cent increase of statistical uncertainty. This problem can also be solved by an additional dataset that provides strong parameter constraints.
Measuring LP from the reconstructed data slightly increases the systematic error but significantly reduces the statistical error, resulting in more accurate measurements.
The 1\,$\sigma$ confidence interval of distance scale constraints from LP measurements are 20--30 per cent larger than those of the corresponding BAO measurements.
For the reconstructed SDSS DR12 data, the constraints on $H_0$ and $\Omega_{\rm m}$ in a flat-$\Lambda$CDM framework with the LP are generally consistent with those from BAO. 
When combined with Planck cosmic microwave background data, we obtain $H_0=68.02_{-0.37}^{+0.36}$ ${\rm km}\,{\rm s}^{-1}\,{\rm Mpc}^{-1}$ and $\Omega_{\rm m}=0.3055_{-0.0048}^{+0.0049}$ with the LP.}
\end{abstract}

\begin{keywords}
methods: data analysis -- cosmological parameters -- distance scale -- large-scale structure of Universe
\end{keywords}



\section{Introduction}
The clustering of large-scale structures (LSSs) of the Universe allows measurements of the cosmic expansion history and structure growth through various physical effects such as baryon acoustic oscillations (BAO) and redshift space distortions (RSD). BAO is a result of primeval acoustic waves propagating in the coupled baryon-photon plasma before decoupling, observed as a peak in the two-point correlation function (2PCF) and as ripples in the power spectrum (PS). It provides us with a powerful standard ruler to map the expansion history of the Universe \citep[][]{Eisenstein1998,Eisenstein2005,Bassett2010}. 
However, there are several physical effects that can shift or distort the peak in the CF, such as nonlinear gravitational evolution \citep[][]{Crocce2008}, RSD and scale dependent bias \citep[][]{Smith2008}.

The {\it Linear Point} (LP), defined as the midpoint between the BAO peak and the associated left dip of the 2PCF, $\xi(s)$, is insensitive to nonlinear gravity and can also be used as a standard ruler \citep[][]{Anselmi2016}. 
The dependences of the LP and the BAO peak position on cosmological parameters are similar \citep[][]{Dwyer2020}.
It has been proved that the differences between the LP positions of linear 2PCF and Zel'dovich-approximated 2PCFs at different redshifts are consistent at 0.5 per cent level \citep[][]{Anselmi2016}. 
In some previous analyses \citep[][]{Anselmi2016, 1711.09063,Anselmi1703}, the LP position of a simulated 2PCF at low redshift is measured using a simple polynomial and then multiplied by 1.005 to restore agreement with linear prediction. \citet[][]{Nikakhtar2021} proposed a Laguerre-based fitting and reconstruction method that provides more reliable LP measurements, especially when the correlation function does not show a clear peak or a dip around the BAO scale. 

This purely geometric (PG) approach drops the reliance on the fiducial power spectrum, which is necessary for the standard template-based BAO analysis. 
It has been identified that LP is independent of the primordial cosmological parameters \citep[][]{Anselmi2016}. 
Therefore, LP is promising for exploring cosmological models beyond LCDM. 
The forecast of $s_{_{\rm LP}}/D_{\rm _V}$ \citep[][]{Anselmi2022}, assuming DESI\footnote{\url{ http://desi.lbl.gov}} \citep[][]{DESI2016} and Euclid\footnote{\url{ http://sci.esa.int/euclid/}} \citep[][]{Euclid2011}, has a little larger uncertainty than standard template-based BAO measured $r_{\rm d}/D_{\rm _V}$, but more accurate than results from {\it Purely-Geometric-BAO} (PG-BAO) method, which is a correlation-function model-fitting (CF-MF) analysis relies on a phenomenological cosmological model for the correlation function. 

The comparison between the $D_{\rm _V}$ derived from LP and BAO position measured from the Baryon Oscillation Spectroscopic Survey \citep[BOSS;][]{Dawson2013} LOWZ and CMASS data shows that the error of the $D_{_{\rm V}}$ from LP analysis is less than that from the standard BAO method with the same pre-reconstruction data, but greater than that of BAO measurement from post-reconstruction data \citep[][]{Anselmi1703}. 
The BAO reconstruction, i.e., the process of eliminating nonlinear effects by reversing motions of galaxies (reference), has never been used in LP analysis because (1) the cosmology of reconstruction, which may deviate from the true cosmology, may bias the measurements \citep[][]{Anselmi1703};
(2) since LP is insensitive to nonlinear effects, reconstruction may not help much. 
However, given that the additional systematic error from BAO reconstruction is less than 0.1 per cent \citep[][]{Carter2020} and reconstruction is able to statistically improve the significance of the BAO peak, reconstruction still be relevant to the LP analysis and deserves careful investigation. 

In this work, we improve the measurements and complete the analysis of LP based on the BOSS LRG dataset and the corresponding approximate mock catalogues. 
We use mocks to explore the performances of LP measurements with reconstruction and investigate the potential systematic biases and reliability of various LP measurement schemes. 
Since the statistical uncertainties are generally much larger than systematic biases for our samples, we use the quintic polynomial for measuring LP as in \citet[][]{1711.09063}, rather than the more advanced method proposed by \citet[][]{Nikakhtar2021} for simplicity.
Furthermore, we perform cosmological parameter constraints for LP measurements from observational data for the first time.

The data and mock catalogues we used in this paper are introduced in Section~\ref{sec:data}. Then Section~\ref{sec:Methodology} presents the methodology for both the LP and BAO measurements. We explore the best LP measurement method and investigate the potential systematic bias using mock catalogues in Section~\ref{sec:Systematic}, and apply our LP analysis to SDSS data and compare with the standard BAO analysis in Section~\ref{sec:results}. The results are concluded in Section~\ref{sec:conclusions}.

\section{Data}
\label{sec:data}

From 2009 to 2014, the BOSS of SDSS--\uppercase\expandafter{\romannumeral 3} \citet[][]{Eisenstein2011} measured over 1.3 million LRG spectra using the double-armed spectrographs \citep[][]{Smee2013} as well as the 2.5-metre Sloan Telescope \citep[][]{Gunn2006} at the Apache Point Observatory. 
According to the different algorithms for targe selection, the BOSS LRG dataset can be divided into LOWZ and CMASS populations, for galaxies with redshifts $z \lesssim 0.4$ and $0.4 < z< 0.7$, respectively. 
As the differences in clustering amplitude of subsamples with different target selection criteria are insignificant \citep[][]{Reid2016, Alam2017}, the LOWZ and CMASS subsamples are combined as final SDSS Data Release 12 (DR12) LRG catalogue for LSS analysis. 
Moreover, BOSS DR12 also includes LRGs from SDSS-\uppercase\expandafter{\romannumeral 1}/\uppercase\expandafter{\romannumeral 2} \citep[][]{Abazajian2009}.
This dataset footprints in both northern and southern galactic caps (NGC and SGC) covering a total of nearly 10000 deg$^2$ sky. Following \citet[][]{Alam2017}, We divide the LRG catalogue into two non-overlapping redshift bins, $0.2< z< 0.5$ and $0.5< z< 0.75$, with the corresponding effective redshifts being 0.38 and 0.61 \citep[e.g.][Section 2.5]{Bautista2021}, respectively. We denote these two subsamples by `low-z' and `high-z' hereafter. 

We use the DR12 MultiDark-Patchy (MD-Patchy) mock catalogues\footnote{\url{https://data.sdss.org/sas/dr12/boss/lss/dr12_multidark_patchy_mocks/}} \citep[][]{Kitaura2016} to estimate covariance matrices for 2PCFs of DR12 LRG data, explore potential systematic bias and performance on cosmological parameter constraints for our LP analysis. 
The cosmological parameters of the MD-Patchy mocks are: $h=0.6777$, $\Omega_{\rm m}= 0.307115$, $\Omega_{\rm b} h^2= 0.02214$, $\sigma_8 = 0.8288$, $n_s = 0.9611$, $\Sigma m_{\nu} = 0\,{\rm eV}$, $r_{\rm d}= 147.66\,{\rm Mpc}$.

The MD-Patchy mocks can reproduce the clustering of the BOSS DR12 data accurately since the \textsc{patchy} code takes advantages of structure formation model based on Augmented Lagrangian Perturbation Theory \citep[ALPT;][]{Kitaura2013} and encodes nonlinear, stochastic, and non-local effects through galaxy biases \citep[][]{Kitaura2014}. 
To investigate the effects of reconstruction on LP measurements, we examine for both pre- and post-reconstruction data. The reconstruction method used in BOSS DR12 galaxies and the corresponding MD-Patchy mocks following the algorithm introduced in \citet[][]{Padmanabhan2012}. The parameters used for reconstruction are $f$=0.757, $b$=1.85, $N_{\rm grid}=512^3$, the smoothing scale $\Sigma_r$ is 15\,$h^{-1}{\rm Mpc}$. 
The NGC and SGC samples use the same parameters, but are processed separately since they are distributed far away from each other. 
In this work, the covariance matrices of pre-reconstruction data are estimated by 2048 independent realizations of DR12 MD-Patchy mocks, while the covariance matrices of post-reconstruction data are estimated by 1000 reconstructed MD-Patchy mocks.

\section{Methodology}
\label{sec:Methodology}
\subsection{Correlation function estimator}
Both the BAO peak and the LP position can be measured from the monopole two-point correlation functions (2PCFs). In this work, we use the Fast Correlation Function Calculator\footnote{\url{https://github.com/cheng-zhao/FCFC}} \citep[\textsc{fcfc};][]{Zhao2023} to calculate the pair counts of the data and random catalogues with the fiducial cosmology: $h=0.6777$, $\Omega_{\rm m}= 0.31$, $\Omega_{\rm b} h^2= 0.02214$, $\sigma_8 = 0.8288$, $n_s = 0.9611$, $\Sigma m_{\nu} = 0.06\,{\rm eV}$.
Then the 2PCFs are estimated through combined pair counts using the Landy--Szalay (LS) estimator \citep[][]{Landy1993}:
\begin{equation}
\xi = \frac{ {\rm DD} - 2 {\rm DR} + {\rm RR}}{{\rm RR}}.
\label{eq:xi}
\end{equation}
Here, the D and R indicate the data and random catalogues, respectively. For the reconstructed galaxy sample, the 2PCF estimator is slightly different, as:
\begin{equation}
\xi = \frac{ {\rm DD} - 2 {\rm DS} + {\rm SS}}{{\rm RR}},
\label{eq:xi_recon}
\end{equation}
in which S stands for the shifted random catalogue, which is generated by moving random objects with the same displacement field for galaxy reconstruction.

Since each mock consists of two datasets corresponding to NGC and SGC, we combine these two datasets following \citet[][]{Zhao2022} at the pair count level. In this work, the 2PCFs we used are all calculated with the combination of NGC and SGC.

\subsection{LP measurements}
\label{subsec:LP_measure}
\subsubsection{LP detection}

The fiducial cosmology used to convert the measured angles and redshifts into comoving coordinates may deviate from the true cosmology, leading to the Alcock--Paczynski (AP) distortion effect. The effect in the 2PCF monopole can be mitigated by rescaling the distance $s$ by the isotropic-volume-distance $D_{_{\rm V}}$ \citep[][]{Ariel2012}. The relation between the true and fiducial CF is:
\begin{equation}
\xi_0(s^{\rm fid}/D_{_{\rm V}}^{\rm fid})\simeq \xi_0(s^{\rm true}/D_{_{\rm V}}^{\rm true}),
\label{eq:CF_relations}
\end{equation}
where
\begin{equation}
D_{_{\rm V}} (z) = \left[ (1+z)^2 D_{_{\rm A}}^2 (z) \frac{cz}{H(z)} \right]^{1/3},
\label{eq:Dv}
\end{equation}
in which $D_{_{\rm A}}$ is the angular diameter distance, $H(z)$ is the Hubble rate, and $c$ is the speed of light.
Following \citet[][]{1711.09063}, the CF can be fitted with a 5$^{\rm th}$-order polynomial,
\begin{equation}
\xi_{\rm fit}(s) = \sum_{i=0}^5 a_i s^i, 
\label{eq:polynomial}
\end{equation}
 with an $s$ bin of $3\,h^{-1}{\rm Mpc}$ and fitting range of 60--130\,$h^{-1}{\rm Mpc}$. The polynomial fits are performed using the Markov Chain Monte Carlo (MCMC) method with \textsc{emcee}\footnote{\url{https://github.com/dfm/emcee}} \citep[][]{emcee}. We consider 32 walkers with 15000 sampling steps each, then remove the first 10000 steps of each walker as burn-in period. 
 
Then the LP position can be calculated by 
\begin{equation}
s_{_{\rm LP}} = \frac{1}{2}(s_{_{\rm peak}}+s_{_{\rm dip}}), 
\label{eq:S-LP-def}
\end{equation}
where $s_{_{\rm peak}}$ and $s_{_{\rm dip}}$ are, respectively, the BAO peak and its associated left dip, which can be obtained by searching for points within the range of $s\in[70, 115]\,h^{-1}{\rm Mpc}$ that satisfy $d\xi_{\rm fit}/ds=0$. The final measurement of LP is given by  
\begin{equation}
y_{_{\rm LP}}=s_{_{\rm LP}}/D_{_{\rm V}}. 
\label{eq:y-def}
\end{equation}
However, $s_{_{\rm LP}}$ defined by Eq.~\eqref{eq:S-LP-def} is not always measurable from the Patchy mocks in practice, as the BAO feature of the 2PCF is not always prominent due to cosmic variances, especially for the pre-reconstruction case. If this happens often, the LP position measurements would be unreliable. Therefore, we also measure the LP positions from rescaled 2PCFs of $s^2 \xi$, which amplifies the significance of the BAO peak, that is, the BAO peak and its associated left dip are calculated by $d(s^2\xi_{\rm fit})/ds=0$. Moreover, BAO reconstruction is foreseen to help settling the problem, as it is expected to statistically improve the significance of the BAO peak. The comparisons between the two LP detection schemes and the measurements based on pre- and post-reconstruction data are detailed in Section~\ref{sec:Systematic}.
\subsubsection{Error estimation}
In \citet[][]{1711.09063}, the uncertainty on the LP position is normally estimated by propagating the uncertainties of the fitted polynomial coefficients (error propagation, EP hereafter), The formula can be written as follows:

\begin{equation}
\sigma_{s_{_{\rm LP}}}= A\cdot C\cdot A^{T},
\end{equation}
in which $C$ is the covariance matrix of the polynomial coefficients from the MCMC sampler, and 
\begin{equation}
A = [\frac{\partial s_{_{\rm LP}}}{\partial a_1}, \frac{\partial s_{_{\rm LP}}}{\partial a_2}, ..., \frac{\partial s_{_{\rm LP}}}{\partial a_n}]\\,
\end{equation}
with
\begin{equation}
 \frac{\partial s_{_{\rm LP}}}{\partial a_i} = \frac{s_{_{\rm LP}}(a_i+\epsilon_i)-s_{_{\rm LP}}(a_i-\epsilon_i)}{2*\epsilon_i},
\label{eq:numerical}
\end{equation}
where $a_i$ indicate the best-fitting polynomial coefficients.
We test $\epsilon_i$ in the range $10^{-10}a_i<\epsilon_i<10^{-4}a_i$, and find that $\sigma_{s_{_{\rm LP}}}$ is stable with $10^{-9}a_i<\epsilon_i<10^{-6}a_i$.
 
In this paper, we also consider another scheme for fitted error estimation, the Bayesian samplers, which is expected to be more promising. 
For fitted polynomial coefficients in Monte--Carlo Bayesian posterior sampling, the posterior distribution of $y_{_{\rm LP}}$ is obtained by calculating $y_{_{\rm LP}}$ as a derived parameter from fitted coefficients. 
And then the $1\,\sigma$ statistical uncertainty can be estimated by mean of 16th and 84th percentage of the posterior distribution of $y_{_{\rm LP}}$. The comparison between Bayesian samplers (MCMC hereafter) and error propagation methods will be described in Section~\ref{sec:Systematic}. 

\subsection{BAO measurements}

The position of the BAO peak can be measured by traditional template fitting \citep[][]{Xu2012}, which can be defined as:
\begin{equation}
\xi_{\rm model} (s) = B^2 \xi_{\rm t} (\alpha s) + A(s),
\end{equation}
where $B$ is normalization factor, $A(s)$ term consists of nuisance parameters accounting for the broad band signal in correlation function with:
\begin{equation}
A(s) = a_0 \, s^{-2} + a_1 \, s^{-1} + a_2, 
\end{equation}
and $\xi_{\rm t}$ indicates the template correlation function:
\begin{equation}
\xi_{\rm t} (s) = \int \frac{k^2\,{\rm d} k}{2 {\pi}^2} \, P_{\rm t} (k) j_0 (k s) \, {\rm e}^{-k^2 a^2} ,
\label{eq:xi_temp}
\end{equation}
which is a Hankel transform of a template power spectrum $P_{\rm t}$, and $j_0$ is the 0-order spherical Bessel function $j_0={\rm sin}(kr)/kr$ , a= 1\,$h^{-1}{\rm Mpc}$ is a parameter to reduce numerical instability of the integration. The template power spectrum $P_{\rm t}$ is defined by:
\begin{equation}
P_{\rm t} (k) = [P_{\rm lin}-P_{\rm smooth}]e^{-k^2\Sigma_{\rm NL}^2/2}+P_{\rm smooth}(k),
\label{eq:PS_temp}
\end{equation}
where $P_{\rm lin}$ is the linear power spectrum, and $P_{\rm smooth}$ is the no-wiggle power spectrum, $\Sigma_{\rm NL}$ is a BAO damping parameter that is used to model the degradation in the acoustic peak due to nonlinear evolution.  

There are 6 parameters ($B^2$, $\alpha$, $\Sigma_{\rm NL}$, $a_1$, $a_2$, $a_3$) in BAO model. The critical parameter $\alpha$ quantifies the dilation of the measured correlation function relative to the template correlation function, it is essentially a measurement of the BAO peak. Our BAO fitting range is 30--180$\,h^{-1}{\rm Mpc}$ with the bin size of 3\,$h^{-1}{\rm Mpc}$. we set flat priors for the parameters ($\alpha$, B, $\Sigma_{\rm NL}$) in the ranges of ([0.8, 1.2], [0, 10], [0, 20])\,$h^{-1}{\rm Mpc}$ , respectively. The priors of the parameters are large enough compared to the posterior distributions.

With the fitted $\alpha$, we can obtain $D_{_{\rm V}}(z)/r_{\rm d}$ based on the fiducial value of the template by
\begin{equation}
D_{_{\rm V}}(z)/r_{\rm d} = \alpha \times D_{\rm _V, fid} / r_{\rm d, fid},
\label{eq:Dv_rd}
\end{equation}
which is used to perform cosmological parameter constraints.

\subsection{Cosmological parameter constraints}

We use the \textsc{cobaya}\footnote{\url{https://github.com/CobayaSampler/cobaya}} package to constrain cosmological parameters with LP and BAO measurements. \textsc{cobaya} is a MCMC sampler implemented in \textsc{python}. In this work, we consider the standard flat-$\Lambda$CDM cosmology, with alternative probes including the Big Bang Nucleosynthesis (BBN) calculations with primordial deuterium abundance \citep[][]{Cooke2018}, which provides the constraints on $r_{\rm d}$, as well as the Planck CMB temperature and polarization data \citep[][]{Planck2020}, to break parameter degeneracies and achieve better cosmological constraints.
For both constraints with LP and BAO, the convergence stop criteria are set into the potential scale reduction factor, $\rm R-1=0.01$. For each chain, we remove the first 30 per cent samples as burn in period.

The likelihood of cosmological parameters $\boldsymbol{p}$ is:
\begin{equation}
\mathcal{L} \approx {\rm e}^{ - \chi^2 (\boldsymbol{p}) / 2 }.
\end{equation}
Since datasets that we used in this paper are independent, here chi-squared function $\chi^2$ can be calculated by
\begin{equation}
\chi^2 (\boldsymbol{p}) = \sum_{z\,{\rm bins}}(y_{_{\rm LP, data}}-y_{_{\rm LP, model}}(\boldsymbol{p}))^2/\sigma_{y}^2,
\label{eq:chi2_y}
\end{equation}
where $y_{_{\rm LP, data}}$ is best-fitting $y_{_{\rm LP}}$ measured from data, and $y_{_{\rm LP, model}}$ is $y_{_{\rm LP}}$ measured from theoretical 2PCFs generated using \textsc{camb}\footnote{\url{https://camb.info/}} software with sampled cosmological parameters. The measuring schemes for $y_{_{\rm LP, data}}$ and $y_{_{\rm LP, model}}$ are the same, for example, if $y_{_{\rm LP, data}}$ is measured from a rescaled 2PCF, $y_{_{\rm LP, model}}$ is also measured from a rescaled 2PCF, but for a theoretical one. 
Note that for theoretical 2PCFs in different cosmology, the searching range of BAO peak and its associated left dip is $s\in$[0.7$r_{\rm d}$, 1.2$r_{\rm d}$]. $r_{\rm d}$ denotes the comoving sound horizon at the drag epoch in the corresponding cosmology. 
The formulae for the BAO measurements are the same as those for the LP measurements, but with $y_{\rm LP}$ replaced by $D_{_{\rm V}} /r_{\rm d}$.

\section{Reliability and error analysis using mock catalogues}
\label{sec:Systematic}
We have introduced some LP measurement schemes which may lead to different best-fitting values and errors of the LP.
In this section, we will take advantage of Patchy mocks to compare these measurement schemes from three aspects: reliability, systematic bias and statistical error, to figure out the optimal scheme for LP analysis. 

\subsection{Reliability}
\label{subsec:Peak_detection}
As we mentioned before, if the BAO feature of the 2PCF is not prominent, $s_{_{\rm LP}}$ would not be measurable. The reliability of LP measurement can be characterized using the proportion of the $s_{_{\rm LP}}$-measurable 2PCFs in all mock realizations.

We show the $s_{_{\rm LP}}$-measurable 2PCF fraction of the best-fitting 2PCFs of the mock realizations, as well as the averaged $\rm N_{peak, MC}/N_{all, MC}$ over all realizations in Table~\ref{tab:all_detect_rate}, where $\rm N_{all, MC}$ refers to the total number of MCMC iterations, and $\rm N_{peak, MC}$ indicates the number of iterations with $s_{_{\rm LP}}$ measurable. 
 
For the pre-reconstruction case, 23--40 per cent of plain 2PCFs are not $s_{_{\rm LP}}$-measurable, indicating that LP measurements from the pre-reconstruction case are unreliable. 
We then tested the reconstruction and 2PCF rescaling that amplify the significance of the BAO peak. With BAO reconstruction, the $S_{_{\rm LP}}$-measurable rate becomes over $\sim$95 per cent, and the rescaling increased the $S_{_{\rm LP}}$-measurable rate to over $\sim$90 per cent. When both operations are used, the $s_{_{\rm LP}}$-measurable rate can be higher than 98 per cent.

\begin{table}
    \centering
    \begin{tabular}{ccccc}
    \hline
    \multirow{3}*{Sample} & \multicolumn{2}{c}{$\xi$} & \multicolumn{2}{c}{$s^2\xi$} \\ \cmidrule(r){2-3} \cmidrule(r){4-5}
      & MCMC    &    best   & MCMC   & best \\
    \cmidrule(r){1-1} \cmidrule(r){2-3} \cmidrule(r){4-5}
    pre-recon (low-z) & 60.6\% & 66.8\% &91.5\% & 98.4\% \\
    pre-recon (high-z) & 66.8\% & 77.3\% &96.1\% & 99.5\% \\
    post-recon (low-z)& 94.9\% & 99.6\% &98.9\% & 100.0\% \\
    post-recon (high-z)& 96.1\% & 99.5\% &99.4\% & 99.9\% \\
    \hline
    \end{tabular}
    \caption{The percentage of $s_{_{\rm LP}}$-measurable 2PCFs of individual mocks. Columns (second and third) labeled with $\xi$ shows measurements from plain 2PCFs, while $s^2\xi$ refers to rescaled 2PCFs. The values in `MCMC' column denoted the mean of $\rm N_{peak, MC}/N_{all, MC}$ of mocks, $\rm N_{all, MC}$ refer to total number of MCMC iterations, and $\rm N_{peak, MC}$ indicates the number of iterations with $s_{_{\rm LP}}$ measurable. The `best' columns show $s_{_{\rm LP}}$-measurable mock fractions.}
    \label{tab:all_detect_rate}
\end{table}

\subsection{Systematic bias}
\label{subsec:systematic}
The systematic biases of $y_{_{\rm LP}}$ are estimated by the difference between the fitted values of mean 2PCFs of mocks and the theoretical prediction. 
To eliminate the potential influence of statistical uncertainty, the covariance measured using the mocks is rescaled by 1/$\rm N_{mocks}$, where $\rm N_{mocks}$ indicates the total number of mocks. 

We show the systematic biases expressed in terms of a percentage in Table~\ref{tab:systematic_all}. The columns under `$\Delta y_{\rm lin}$' show $\Delta y/y_{_{\rm LP, lin}}$ with $\Delta y = y_{_{\rm LP, fit}}-y_{_{\rm LP, lin}}$, in which $y_{_{\rm LP, lin}}$ is $y_{_{\rm LP}}$ predicted by linear theory, and $y_{_{\rm LP, fit}}$ is the best-fitting $y_{_{\rm LP}}$ of mean 2PCFs of mocks.
Best-fitting $y_{_{\rm LP}}$ can be estimated either as the median value of posterior distribution of $y_{_{\rm LP}}$ or as the value corresponding to the minimum $\chi^2$, denoted as `best-fitting (median)' and `best-fitting (min)', respectively. 
For the measurements from the rescaled 2PCFs, both the best-fitting and the theoretical predicted $y_{_{\rm LP}}$ are measured from the rescaled 2PCFs.
It shows that systematic biases are all greater than 1.04 per cent.

\begin{table*}
    \centering
    \begin{tabular}{ccccccccc}
    \hline 
    \multirow{3}*{Sample} & \multicolumn{4}{c}{$\xi$} & \multicolumn{4}{c}{$s^2\xi$} \\ \cmidrule(r){2-5} \cmidrule(r){6-9}
    
    & \multicolumn{2}{c}{$\Delta y_{\rm lin}$} &\multicolumn{2}{c}{$\Delta y_{\rm zel}$}& \multicolumn{2}{c}{$\Delta y_{\rm lin}$} & \multicolumn{2}{c}{$\Delta y_{\rm zel}$}\\
    \cmidrule(r){2-3} \cmidrule(r){4-5} \cmidrule(r){6-7} \cmidrule(r){8-9}
                     & med & min & med & min & med & min  & med & min\\
    \cmidrule(r){1-1} \cmidrule(r){2-3} \cmidrule(r){4-5} \cmidrule(r){6-7} \cmidrule(r){8-9}
    pre-recon (low-z) & 1.35\% & 1.43\% & 0.28\% & 0.21\%  & 1.59\% & 1.66\% & 0.04\% & 0.03\%  \\
    pre-recon (high-z) & 1.38\% & 1.32\% & 0.26\% & 0.31\%  & 1.57\% & 1.48\% & 0.06\% & 0.15\%  \\
    post-recon (low-z) & 1.43\% & 1.30\% & 0.21\% & 0.33\%  & 1.17\% & 1.04\% & 0.47\% & 0.60\%  \\
    post-recon (high-z) & 1.43\% & 1.38\% & 0.20\% & 0.25\%  & 1.22\% & 1.16\% & 0.42\% & 0.48\%  \\
    \hline
    \end{tabular}
    \caption{The potential systematic bias estimated by mean 2PCF of Patchy mocks. The columns under `$\Delta y_{\rm lin}$' show $(y_{_{\rm LP, fit}}-y_{_{\rm LP, lin}})/y_{_{\rm LP, lin}}$ with $y_{_{\rm LP, lin}}$ being $y_{_{\rm LP}}$ predicted by linear theory, and $y_{_{\rm LP, fit}}$ is the best-fitting $y_{_{\rm LP}}$ of mean 2PCFs of mocks, which can be estimated by the median value of the posterior distribution of $y_{_{\rm LP}}$ (shown in `med' column) and $y_{_{\rm LP}}$ value corresponding to minimum reduced $\chi^2$ (shown in `min' column). $\Delta y_{\rm zel}$ is similar to $\Delta y_{\rm lin}$, but replacing $y_{_{\rm LP, lin}}$ with $y_{_{\rm LP, zel}}$ estimated from the Zel'dovich-approximated 2PCFs. }
    \label{tab:systematic_all}
\end{table*}

Since nonlinear effects can smooth and shift the BAO peak in CF, resulting in  differences between the LP position predicted by linear theory and the measured one. We check also the Zel'dovich-approximated theoretical 2PCFs with the results from mean 2PCFs of mocks.
The Zel'dovich-approximated 2PCFs can be calculated as 
\begin{equation}
\xi^{\rm zel}(s)=\int \frac{dk}{k} \frac{k^3P_{\rm lin}(k)}{2\pi^2} e^{- k^{2} \sigma_{v}^{2}(z)}j_0(ks),
\label{eq:xi_zel}
\end{equation}
where 
\begin{equation}
\sigma_{v}^{2}=\frac{1}{3} \int\frac{d^3q}{(2\pi)^3} \frac{P_{\rm lin}(q, z)}{q^2} 
\label{eq:sigma_v}
\end{equation}
is the square of linear displacement field dispersion and the damping scale $k_{NL}=1/\sigma_{v}$. 

Then Zel'dovich-approximated $y_{_{\rm LP, zel}}$ can be estimated from $\xi^{\rm zel}$. 
We show the biases estimated by $(y_{_{\rm LP, fit}}-y_{_{\rm LP, zel}})/y_{_{\rm LP, zel}}$ in Table~\ref{tab:systematic_all}. 
We find that $y_{_{\rm LP, fit}}$ are significantly more consistent with $y_{_{\rm LP, zel}}$ than $y_{_{\rm LP, lin}}$, and the biases are less than 0.6 per cent for all cases. 
BAO reconstruction slightly increases the systematic bias.
In this case, the median value of $y_{_{\rm LP}}$ posterior distribution is closer to the theoretically predicted $y_{_{\rm LP, zel}}$ than the corresponding minimum-$\chi^2$ value. 

The measurements from the rescaled post-reconstruction 2PCFs show the largest $\Delta y_{\rm zel}$, and the lowest $\Delta y_{\rm lin}$, which may be due to the mitigation of nonlinear effects by reconstruction.
However, given the fact that $y_{_{\rm LP, zel}}$ is more consistent with $y_{_{\rm LP}}$ measured from mean 2PCFs of mocks, the Zel'dovich approximation 2PCFs are adopted as theoretical 2PCFs of $y_{_{\rm LP, model}}$ (Eq.~\ref{eq:chi2_y}) for cosmological parameter constraints, and $\Delta y_{\rm zel}$ will be included in the error estimation.

We also explore the potential systematic error with the fiducial cosmological parameter $\Omega_{\rm m}=\{0.12, 0.25, 0.31, 0.64, 0.85\}$, showing that the measured LPs are still consistent with the theoretically predicted values ($y_{_{\rm LP, zel}}$) at 0.6 per cent level (see Appendix~\ref{sec:sys_oms}). 

Note that the systematic error of $y_{_{\rm LP}}$ can be further reduced by measuring LP using the Laguerre reconstruction method \citep[][]{Nikakhtar2021}; we leave the relevant studies for future work.

\subsection{Statistical uncertainty}
\label{sec:stat_error}
The statistical uncertainty is crucial for cosmological constraints with LP, therefore it is important to identify the LP measurement scheme that provides the most reliable statistical uncertainty.  
In this subsection, we compare the statistical errors of $s_{_{\rm LP}}$ measured with different schemes. 
We show the 1\,$\sigma$ statistical errors of $s_{_{\rm LP}}$ ($\sigma_{s_{_{\rm LP}}}$) estimated from the mean 2PCFs of the mocks and the 2PCFs of the individual mocks in Table~\ref{tab:statitic_all}. 
\begin{table*}
    \centering
    \begin{tabular}{ccccccccccccccc}
    \hline 
    \multirow{3}*{Sample} & \multicolumn{7}{c}{$\xi$} & \multicolumn{7}{c}{$s^2\xi$} \\ \cmidrule(r){2-8} \cmidrule(r){9-15}
    
    & \multicolumn{3}{c}{mean of mocks} &\multicolumn{4}{c}{individual mocks}& \multicolumn{3}{c}{mean of mocks} &\multicolumn{4}{c}{individual mocks}\\
     \cmidrule(r){2-4} \cmidrule(r){5-8} \cmidrule(r){9-11} \cmidrule(r){12-15}
                     & $\sigma_{\rm _{CR}}$& $\sigma_{\rm _{MC}}$ & $\sigma_{\rm _{EP}}$ & $\sigma_{\rm med}$ &  $\sigma_{\rm min}$ &  $\Tilde{\sigma}_{\rm MC}$ &$\Tilde{\sigma}_{\rm EP}$  & $\sigma_{\rm _{CR}}$& $\sigma_{\rm _{MC}}$ & $\sigma_{\rm _{EP}}$ & $\sigma_{\rm med}$ &  $\sigma_{\rm min}$ &  $\Tilde{\sigma}_{\rm MC}$ &$\Tilde{\sigma}_{\rm EP}$ \\
    \cmidrule(r){1-1}\cmidrule(r){2-2} \cmidrule(r){3-4} \cmidrule(r){5-6} \cmidrule(r){7-8} \cmidrule(r){9-9} \cmidrule(r){10-11} \cmidrule(r){12-13}\cmidrule(r){14-15}
    pre (low-z) &2.21 & 2.05 & 2.36 & 2.07 & 1.85 & 1.99 &1.78 &2.27 & 2.24 & 2.03 & 2.18 & 2.10 & 2.17 &2.06 \\
    pre (high-z) &1.91 & 1.81 & 1.56 & 1.89 & 1.69 & 1.78 &1.64 &1.95 & 1.96 & 2.99 & 1.96 & 1.83 & 1.92 &1.81 \\
    post (low-z) &1.46 & 1.47 & 1.43 & 1.47 & 1.32 & 1.48 &1.40 &1.52 & 1.52 & 1.62 & 1.50 & 1.40 & 1.53 &1.51 \\
    post (high-z) &1.33 & 1.36 & 1.12 & 1.32 & 1.26 & 1.36 &1.30 &1.37 & 1.38 & 1.21 & 1.34 & 1.29 & 1.38 &1.33 \\
    \hline 
    \end{tabular}
    \caption{The statistical errors of $s_{_{\rm LP}}$. The columns label with `mean of mocks' show the mean of the lower and upper 1\,$\sigma$ confidence limits of $s_{_{\rm LP}}$ fitted from the mean 2PCFs of mocks. Among them, $\sigma_{\rm _{CR}}$ indicates errors measured from fittings with the covariances rescaled by 1/$\rm N_{mocks}$ and then the 1\,$sigma$ confidence intervals multiplied by $\sqrt{\rm N_{mocks}}$. The $\sigma_{\rm _{MC}}$ and $\sigma_{\rm _{EP}}$ indicate errors measured with MCMC and error propagation method respectively. Notes that except the `$\sigma_{\rm _{CR}}$' column, all of the statistical error show in this table are measured with the covariances are not rescaled. Columns labeled with `individual mocks' show measurements from individual mocks, $\sigma_{\rm med}$ and $\sigma_{\rm min}$ are the $1\,\sigma$ dispersion of best-fitting (median) and best-fitting (min) ${s_{_{\rm LP}}}$ distribution of all mocks, respectively. $\Tilde{\sigma}_{\rm MC}$ and $\Tilde{\sigma}_{\rm EP}$ are the median values of the fitted errors $\sigma_{s_{_{\rm LP}}}$ of individual mocks estimated by the MCMC method and the EP method, respectively.}
    \label{tab:statitic_all}
\end{table*}

 To set a solid reference for the comparison, we rescale the covariance matrix by $1/{\rm N_{mocks}}$ when fitting to the mean 2PCFs of mocks, thus the cosmic variance is highly reduced. Then, we rescale the fitted error of the LP by $\sqrt{N_{\rm mocks}}$, so the value is directly comparable to that from individual mocks.

Comparing to the 1\,$\sigma$ dispersion of best-fitting (min) $s_{_{\rm LP}}$ distribution of all mocks ($\sigma_{\rm min}$), the measurements from best-fitting (median) $s_{_{\rm LP}}$, i.e., $\sigma_{\rm med}$ are more consistent with $\sigma_{\rm _{CR}}$, 
suggesting that the median value is a better indicator of the best-fitting value.  

The statistical errors measured by MCMC method, including measurements from mean 2PCFs of mocks ($\sigma_{\rm _{MC}}$) and median value of $\sigma_{s_{_{\rm LP}}}$ from individual mocks ($\Tilde{\sigma}_{\rm MC}$), are consistent with each other, and generally more consistent with $\sigma_{\rm med}$ than corresponding measurements from the EP method.
Moreover, the EP method also shows higher systematic error. Thus, MCMC is a more reliable method for statistical uncertainty estimation. 

\begin{figure}
\centering
\includegraphics[width=0.98\hsize]{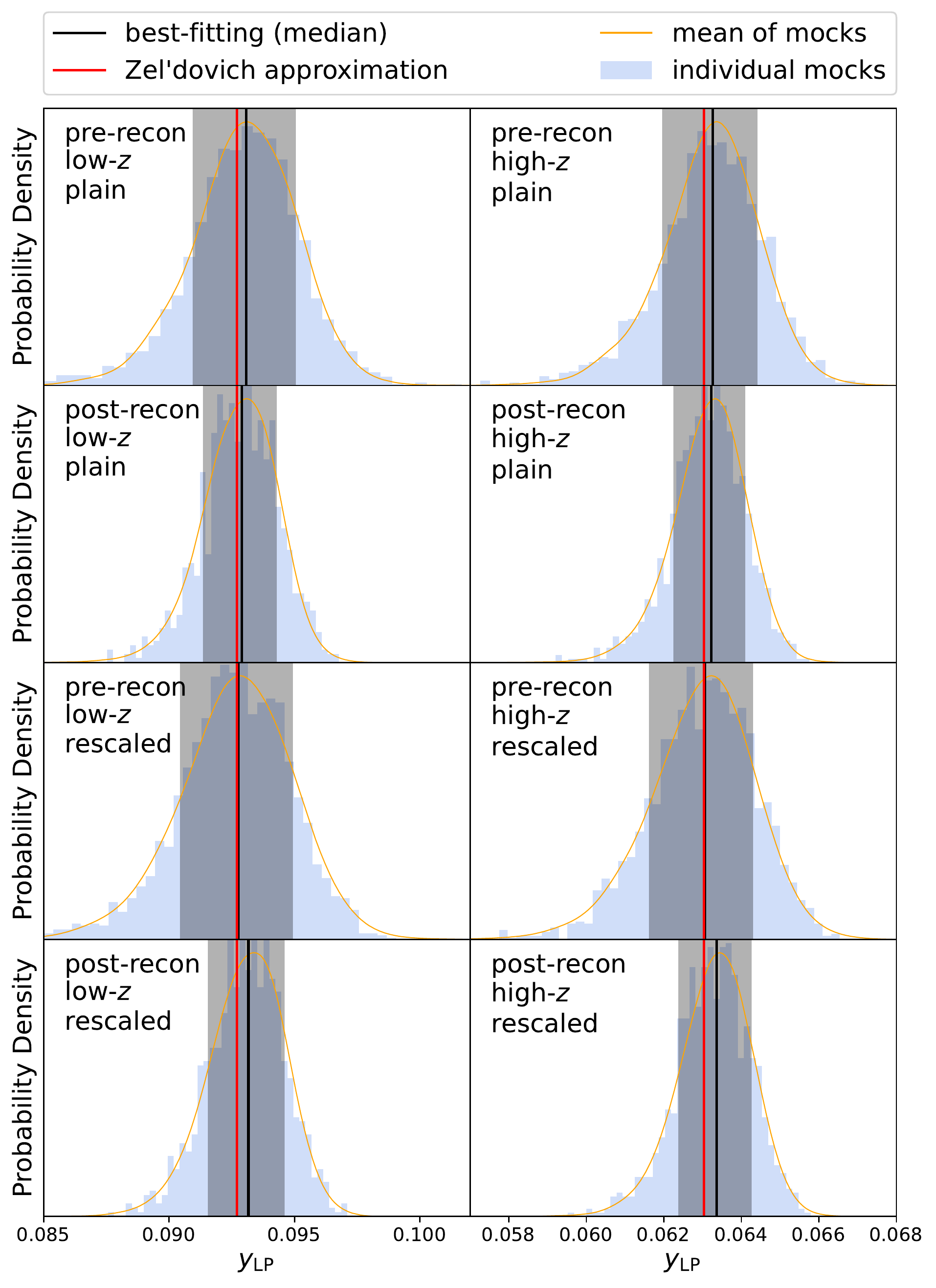}

\caption{The fitted posterior distributions of $y_{_{\rm LP}}$ for the mean 2PCFs of all mocks (orange lines) and distributions of best-fitting (median) $y_{_{\rm LP}}$ of individual mocks (blue histograms). The black vertical line and grey shadows indicate the 50th and range of 16th to 84th percentiles of the cumulative posterior distribution of $y_{_{\rm LP}}$ for the mean 2PCFs of all mocks, respectively. The red vertical lines show the expected $y_{_{\rm LP}}$ from theoretical Zel'dovich-approximated 2PCFs. The upper two rows show two redshift bins of pre-reconstruction, while the bottom two rows show the results of post-reconstruction}
\label{fig:y_dis}
\end{figure}

We show the fitted posterior distributions of $y_{_{\rm LP}}$ from the mean 2PCFs of the mocks (orange lines) and the distribution of best-fitting (median) $y_{_{\rm LP}}$ of individual mocks (blue histograms) in Figure~\ref{fig:y_dis}. 
The posterior of $y_{_{\rm LP}}$ fitted from mean 2PCFs of mocks show good agreement with the distributions of mock realizations for both measurements from plain (upper two rows) and rescaled 2PCFs (bottom two rows), suggesting that the posteriors of $y_{_{\rm LP}}$ are unbiased and the values derived from the posteriors are reliable.

The comparison of statistical errors from the plain and rescaled 2PCFs shows that measuring the LP position from the rescaled 2PCFs increases the statistical error by 2--9 per cent without reconstruction and by 2--4 per cent for the post-reconstruction case. Note however that the measurements from pre-reconstruction plain 2PCFs may not be reliable due to the low $s_{_{\rm LP}}$-measurable rate. 
Comparing the results on the rescaled 2PCFs with and without BAO reconstruction, we can find that reconstruction significantly reduces the statistical uncertainty by 20--30 per cent. 

In summary, statistical errors estimated by EP method are less stable than MCMC method and best-fitting (median) outperforms best-fitting (min). Therefore, we use best-fitting value given by median value of posterior distribution of $y_{_{\rm LP}}$, and the statistical error measured by the MCMC method for the following LP measurements.

\subsection{LP and BAO fitting results}

\begin{figure}
\centering
\includegraphics[width=0.98\columnwidth]{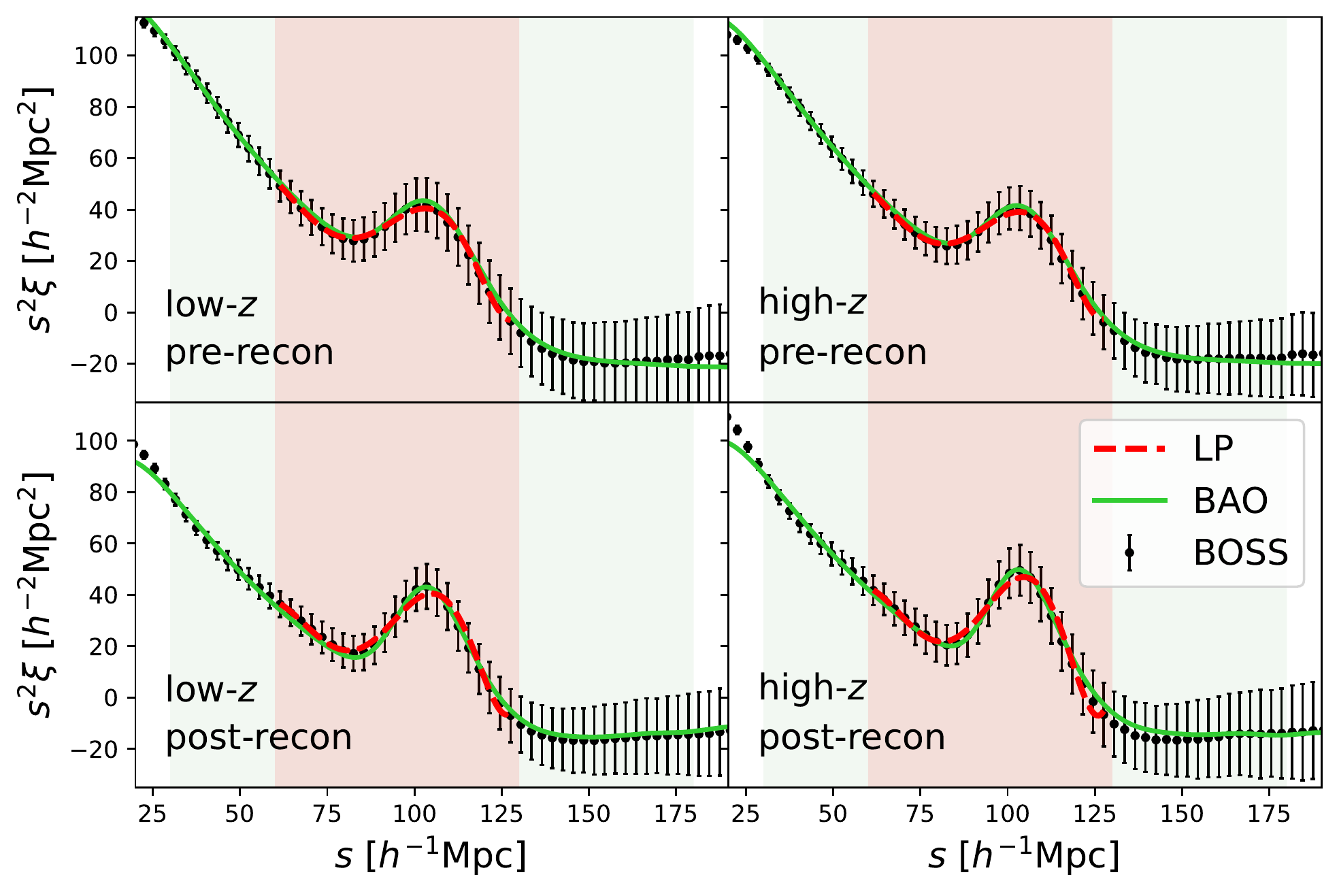}
\caption{The mean 2PCFs of all Patchy mocks of different samples. The dots show the mean 2PCFs of all Patchy mocks, with error bars being the standard deviation of measurements from realizations of the corresponding Patchy mocks. The dashed red lines show the best-fitting 5$^{\rm th}$-order polynomial models. The best-fitting 2PCFs from BAO template fitting are shown with solid green lines. The orange shadow areas are the fitting ranges of polynomial fitting (60--130 $h^{-1}{\rm Mpc}$) and the green ranges are that of BAO fitting (30--180 $h^{-1}{\rm Mpc}$). The upper panels show pre-reconstruction measurements, while bottom panels show post-reconstruction's. }
\label{fig:2PCF_mock}

\end{figure}

\begin{figure}
\centering
\includegraphics[width=1.\hsize]{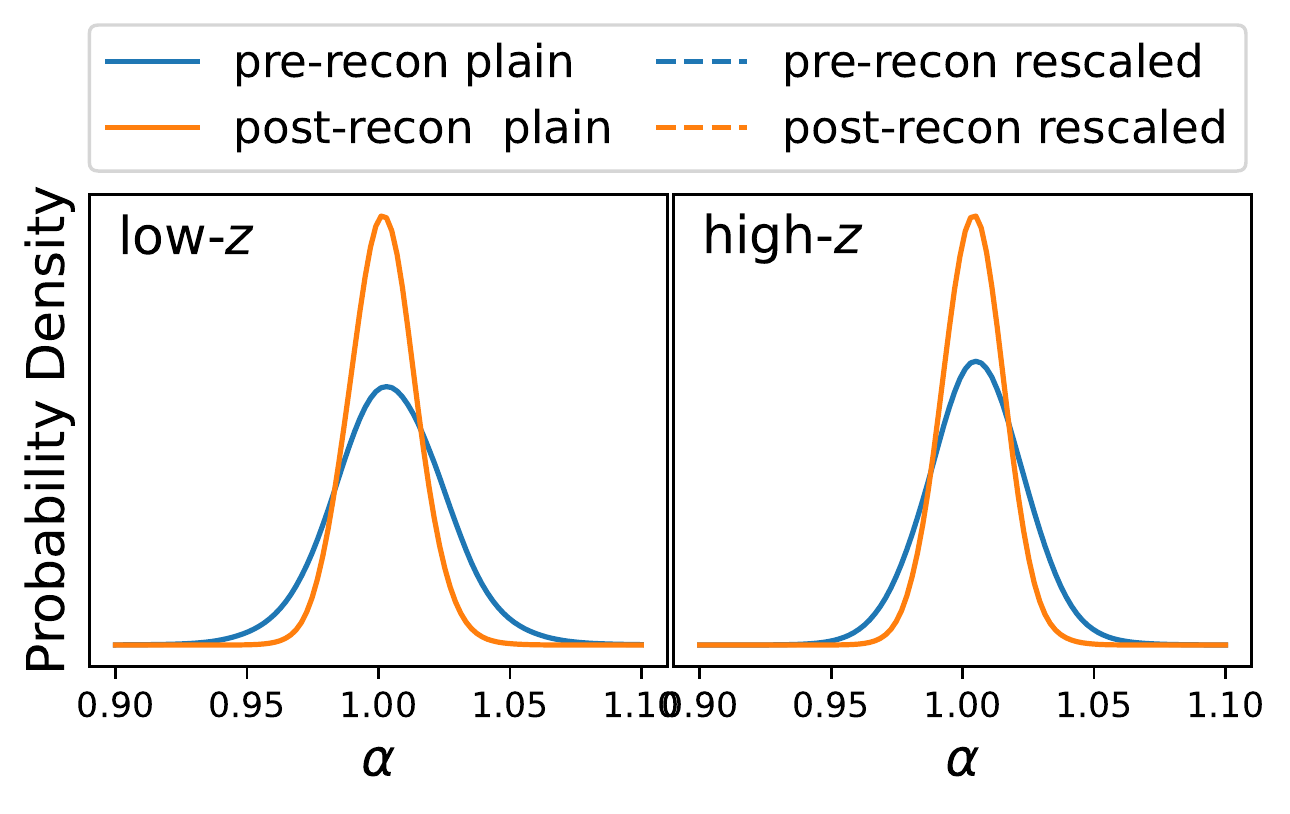}
\includegraphics[width=1.\hsize]{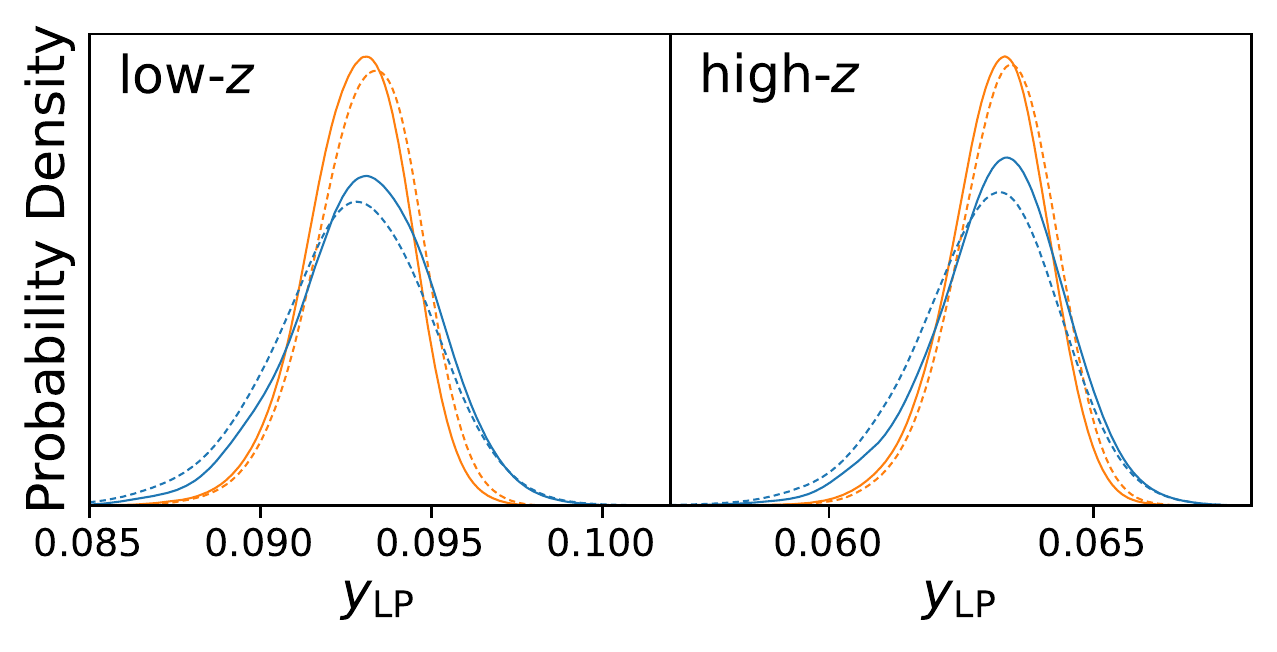}
\caption{The marginalized posterior distribution of $\alpha$ and $y_{_{\rm LP}}$ fitted from the mean 2PCFs of the mocks. The blue lines show measurements from pre-reconstruction data, while measurements from post-reconstruction data are shown with orange lines. The distribution of $y_{_{\rm LP}}$ measured from SDSS data are illustrated in bottom two panels, the dashed lines show distributions of $y_{_{\rm LP}}$ that measured from rescaled 2PCFs, while the solid lines indicate distributions of $y_{_{\rm LP}}$ that measured from plain 2PCFs.}
\label{fig:posterior_mock}
\end{figure}

We compare the LP and BAO measurements from mean 2PCFs of Patchy mocks. The resulting 2PCFs are shown in Figure~\ref{fig:2PCF_mock} with the corresponding posterior distributions of the parameters $y_{_{\rm LP}}$ and $\alpha$ illustrated in Figure~\ref{fig:posterior_mock}. 
The fitted 2PCFs generally agree well with the mean 2PCFs of mocks in their fitting range, but the fitted 2PCFs described by the 5$^{\rm th}$-order polynomial show larger deviations than the 2PCFs fitted with BAO template fitting, especially at the large end of the fitting range.
Figure~\ref{fig:posterior_mock} shows that the distributions of the post-reconstruction data (orange and yellow) are more concentrated than those without reconstruction, especially for the redshift bin $z\in [0.2, 0.5]$, and the distributions of $y_{_{\rm LP}}$ measured from plain 2PCFs are slightly more concentrated than the measurement from rescaled 2PCFs, which is consistent with our findings in the last subsection.


\begin{figure}
\centering
\includegraphics[width=1\hsize]{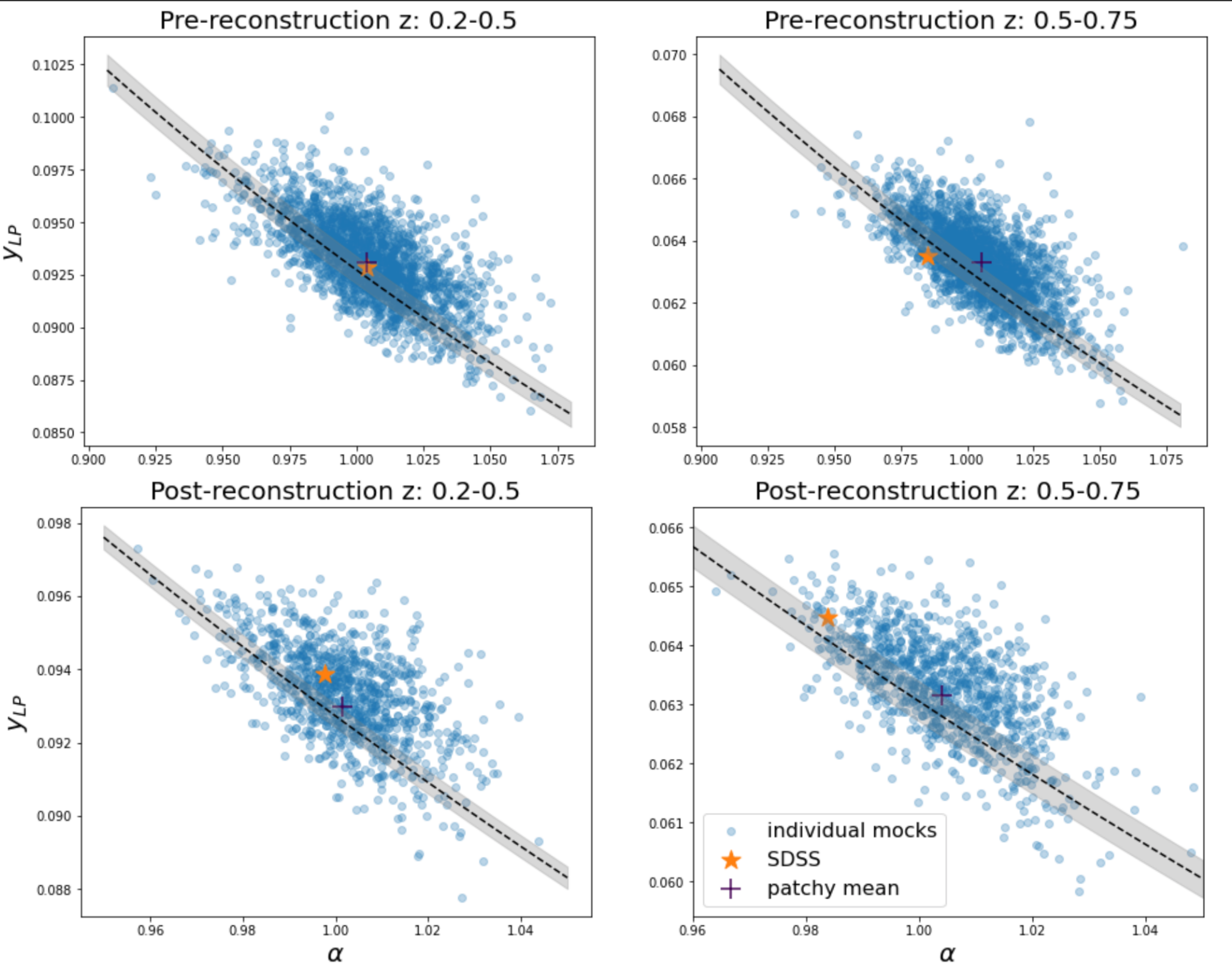}
\includegraphics[width=1\hsize]{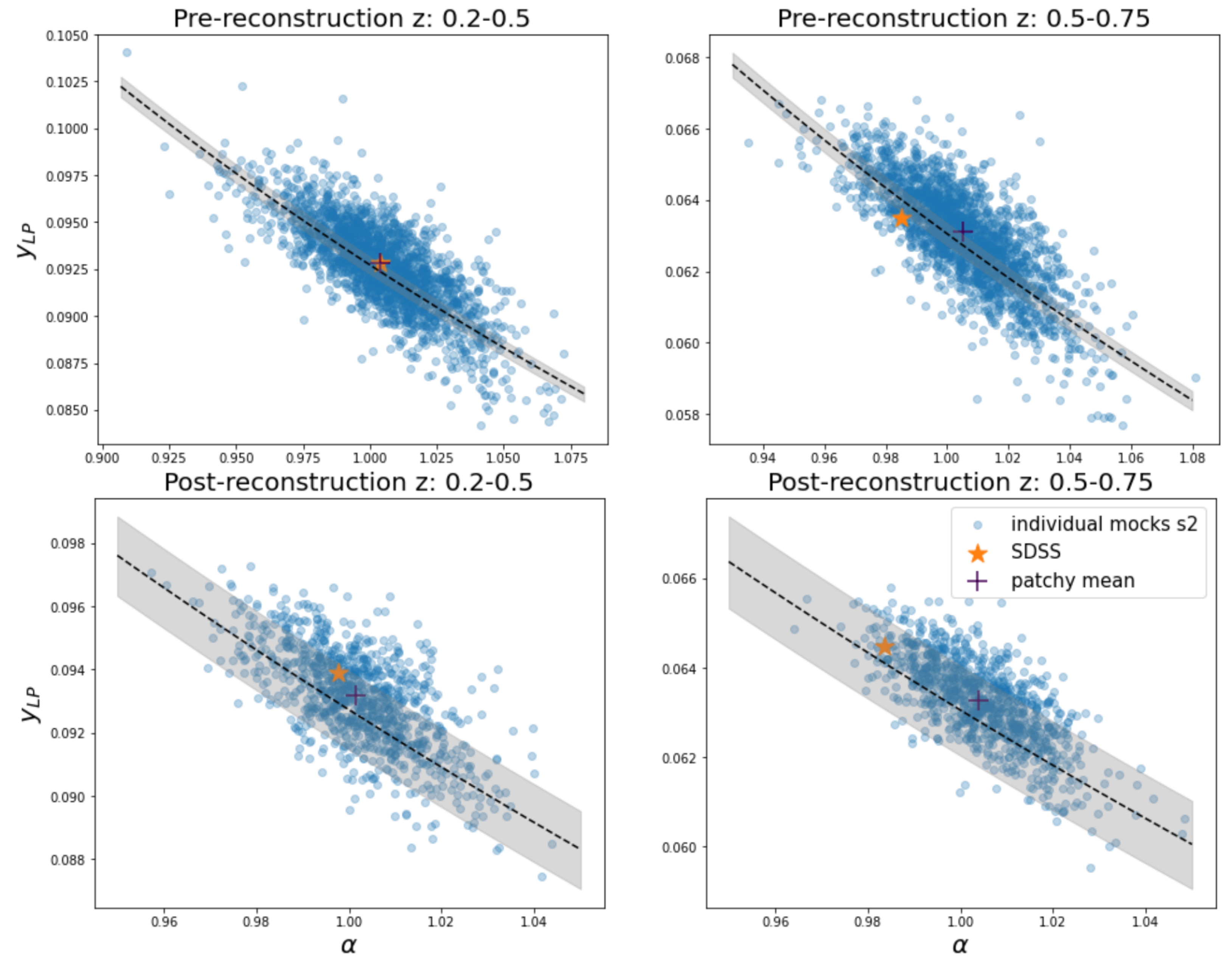}
\caption{Distribution of the best-fitting value of $y_{_{\rm LP}}$ and $\alpha$. The upper two rows show the $y_{_{\rm LP}}$ estimated from plain 2PCfs while the rest two rows show measurements from rescaled 2PCFs. The blue scatter points illustrate the measurements from individual mocks. Purple plus symbols shows best-fitting values from the mean 2PCFs of Patchy mocks, while the yellow star symbols indicate the best-fitting values from SDSS data. The black dashed lines and grey shadow areas show the theoretical predicted values defined as Eq.~\ref{eq:theory_y_alpha} and the corresponding systematic bias range. }
\label{fig:Scatter}
\end{figure}

We also plot the distribution of best-fitting values of $y_{_{\rm LP}}$ and $\alpha$ in Figure~\ref{fig:Scatter}. The black dashed lines indicate the theoretical predicted $y_{_{\rm LP}}$ as a function of $\alpha$, which is calculated by:
\begin{equation}
    y= \frac{1}{(D_{_{\rm V}}/s_{_{\rm LP}})_{\rm true}\times \alpha},
\label{eq:theory_y_alpha}
\end{equation}
where $(D_{_{\rm V}}/s_{_{\rm LP}})_{\rm true}$ is calculated by \textsc{camb} with the cosmology of Patchy mocks, the grey shadow areas denote 1\,$\sigma$ statistical error of $y_{_{\rm LP}}$ and $\alpha$. The systematic error of $y_{_{\rm LP}}$ is computed using $y_{_{\rm LP, zel}}$, as discussed in Section\ref{sec:Systematic}, and the systematic error of $\alpha$ is defined as:
\begin{equation}
    \Delta \alpha = \alpha_{_{\rm best}} -\alpha_{_{\rm exp}},
\label{eq:alpha_sys}
\end{equation}
where $\alpha_{_{\rm exp}}$ is estimated by
\begin{equation}
    \alpha_{_{\rm exp}} = \frac{D^{\rm true}_{\rm _V} / r^{\rm true}_{\rm d}}{D^{\rm fid}_{\rm _V} / r^{\rm fid}_{\rm d}},
\label{eq:alpha_expect}
\end{equation}
and $\alpha_{_{\rm best}}$ denotes the best-fitting $\alpha$ of mean 2PCF of mocks with the covariance matrix rescaled by $1/{\rm N_{mocks}}$.

Figure~\ref{fig:Scatter} shows that $\alpha$ and $y_{_{\rm LP}}$ are highly anti-correlated.
The values from individual mocks are distributed around the theoretical predicted line, so do the measurements from data and mean of all Patchy mocks. It is consistent with the theoretical predictions after taking into account the systematic errors discussed in Section \ref{sec:Systematic}.

\begin{table*}
    \centering
    \begin{tabular}{cccccccc}
    \hline
    Patchy mock & measurement & best-fitting & $\Delta$ & $\sigma$ & $\sigma_{\rm comb}$ &$\sigma_{\rm comb}$(\%) & $\frac{\chi^2}{\rm d.o.f}$ \\
        \hline
        \multirow{3}*{pre  (low-z)} & $ 100\times y_{_{\rm LP}}$($\xi$)&  9.308 & 0.026& 0.206 & 0.208 &  2.2\%& \multirow{2}*{3.55}\\
        & $ 100\times y_{_{\rm LP}}$($s^2\xi$)& 9.276 & 0.004& 0.225 & 0.225 &  2.4\% \\
        \cmidrule(r){2-8}
        & $\alpha$ & 1.0038 & 0.0045& 0.021 & 0.021 &  2.1\% & 0.61\\
        \hline
        \multirow{3}*{pre  (high-z)} & $ 100\times y_{_{\rm LP}}$($\xi$)&  6.327 & 0.016& 0.123 & 0.124 &  2.0\%& \multirow{2}*{3.51}\\
        & $ 100\times y_{_{\rm LP}}$($s^2\xi$)& 6.306 & 0.004& 0.134 & 0.134 &  2.1\%\\
        \cmidrule(r){2-8}
        &  $\alpha$ & 1.0052 & 0.0042& 0.018 & 0.019 &  1.9\% & 0.63\\
        \hline
        \multirow{3}*{post  (low-z)}& LP($\xi$)& 9.289 & 0.019& 0.147 & 0.148 &  1.6\% & \multirow{2}*{9.20}\\
        & $ 100\times y_{_{\rm LP}}$($s^2\xi$)&  9.316 & 0.044& 0.153 & 0.159 &  1.7\% \\
        \cmidrule(r){2-8}
        & $\alpha$ & 1.0016 & 0.0015& 0.013 & 0.013 &  1.3\% & 1.47\\
        \hline
        \multirow{3}*{post  (high-z)}& $ 100\times y_{_{\rm LP}}$($\xi$)&  6.322 & 0.013& 0.093 & 0.094 &  1.5\% & \multirow{2}*{8.00}\\
        & $ 100\times y_{_{\rm LP}}$($s^2\xi$)&  6.336 & 0.026& 0.094 & 0.098 &  1.5\% \\
        \cmidrule(r){2-8}
        &  $\alpha$ & 1.0040 & 0.0036& 0.012 & 0.012 &  1.2\% & 1.59\\
        \hline
    \end{tabular}
    \caption{The measured $y_{_{\rm LP}}$ and $\alpha$ from the mean 2PCFs of the mocks. Note that all values of the $y_{_{\rm LP}}$ measurements, except `$\sigma_{\rm comb}$(\%)', are multiplied by 100. $\Delta$ column shows systematic biases measured from the fittings to the mean of patchy mock with covariances rescaled by $\rm 1/N_{mock}$. $\sigma$ and $\sigma_{\rm comb}$ columns show the statistical errors and combined error $\sigma_{\rm comb} = \sqrt{\Delta^2+\sigma^2}$, respectively. $\frac{\chi^2}{\rm d.o.f}$ shows the best-fit reduced $\chi^2$ of corresponding models, note that $\chi^2$ are estimated with normalized covariance matrices and ${\rm d.o.f}$ indicates degree of freedom. }
    \label{tab:fit_res_mock}
\end{table*}

The measurements of $y_{_{\rm LP}}$ and $\alpha$ are summarised in Table~\ref{tab:fit_res_mock}, in which $\sigma_{\rm comb}= \sqrt{{\Delta}^2+\sigma^2}$ indicates the combined error of systematic bias ${\Delta}$ and statistical error $\sigma$,  which is used for cosmological analysis.
The combined error of $y_{_{\rm LP}}$ is dominated by the statistical error, since the potential systematic bias (difference between the black and red vertical lines in Figure~\ref{fig:y_dis}) is much less than the statistical error (shadow range in Figure~\ref{fig:y_dis}). 
For $y_{_{\rm LP}}$ measurement, BAO reconstruction is able to reduce the combined error by 25--32 per cent.
Though with the rescaled 2PCF the LP measurements are more reliable, the statistical errors become 5--8 per cent larger. 
Comparing the post-reconstruction results with the corresponding $\alpha$ measurements, the 1\,$\sigma$ confidence interval of $y_{_{\rm LP}}$ is 20--30 per cent larger.

\subsection{Cosmological parameter measurements}
\label{sec:mock_Cospar}

We convert the best-fitting $\alpha$ to $D_{_{\rm V}} / r_{\rm d}$ measurements, and then perform cosmological parameter constraints in the standard flat-$\Lambda$CDM cosmological model with different $D_{_{\rm V}} / r_{\rm d}$ measurements and $y_{_{\rm LP}}$ measurements from mean 2PCFs of mocks. 

\begin{figure*}
\centering
\includegraphics[width=0.8\columnwidth]{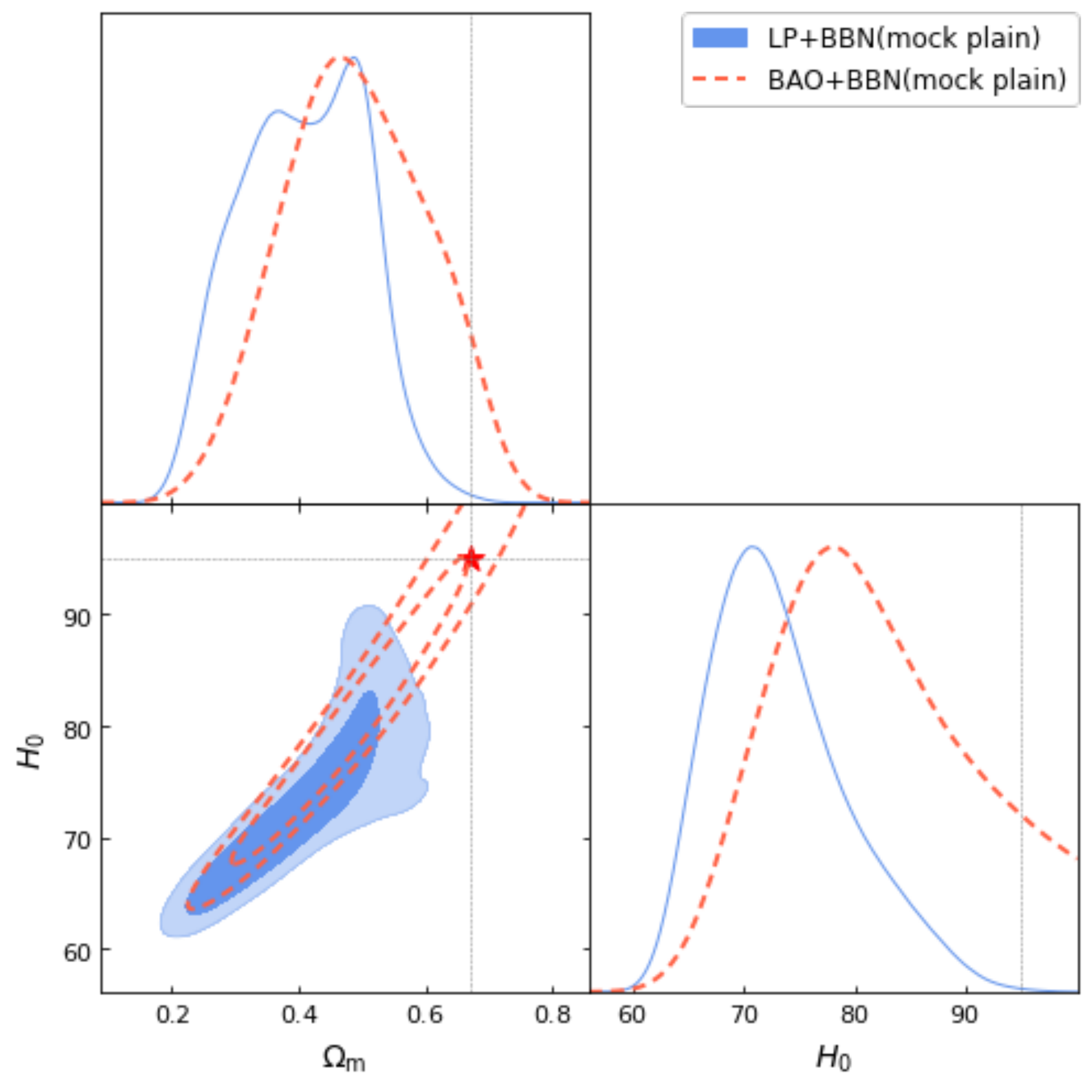}
\includegraphics[width=0.8\columnwidth]{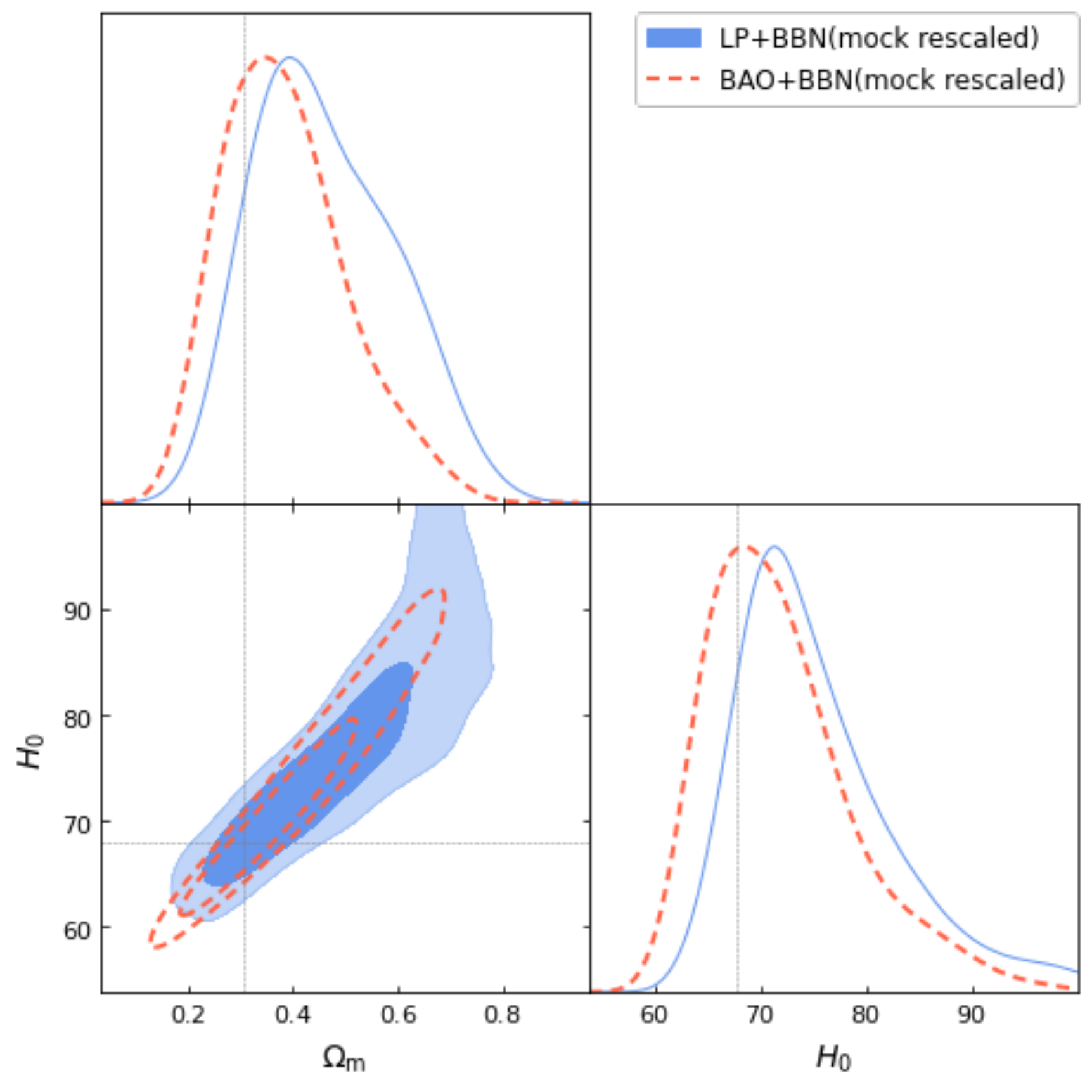}
\caption{Constraints on $H_0$ and $\Omega_{\rm m}$ based on combination of BBN and measurements from reconstructed mocks. The blue contour and lines show constraints from BBN + LP, while the red dashed contour and lines show that of BBN + BAO. The left panel shows result of LP measured from plain 2PCFs and the right panel is for LP from rescaled 2PCFs. 
Notes that, results of BBN + BAO are the same in left and right panels. The grey dotted lines indicate cosmological parameters used for mock construction. The contours show the 68 (1\,$\sigma$) and 95 per cent (2\,$\sigma$) confidence intervals. The red star indicates a parameter set at the 1\,$\sigma$ edge of the BAO constraints, but excluded by the LP measurements at more than 2\,$\sigma$ level. }
\label{fig:cospar_mock}
\end{figure*}
\begin{table}
    \centering
    \begin{tabular}{cccc}
    \hline
                  & true     &  BAO                 &  LP \\
         \hline
        $H_0$ (${\rm km}\,{\rm s}^{-1}\,{\rm Mpc}^{-1}$) & 67.77    & $71.4_{-7.8}^{+4.4}$   & $75.3_{-8.7}^{+4.4}$\\
        \\
        $\Omega_{\rm m}$& 0.307115 &$0.376_{-0.131}^{+0.091}$&$0.459_{-0.141}^{+0.146}$\\
    \hline
    \end{tabular}

    \caption{The best-fitting cosmological parameters and 1\,$\sigma$ confidence intervals of LP and BAO measured from mean 2PCFs of Patchy mocks. The cosmological parameters used for mock construction are shown in 'true' column. }
    \label{tab:cospar_mock}
\end{table}

The constraints of $H_0$ and $\Omega_{\rm m}$ from the combination of the BBN and $y_{_{\rm LP}}$ measurements (BBN + LP) and the combination of the BBN and $D_{_{\rm V}} / r_{\rm d}$ measurements (BBN + BAO) are shown in Figure~\ref{fig:cospar_mock} and Table~\ref{tab:cospar_mock}. Note that all measurements are from reconstructed data.
Compared to the constraints from the BBN + LP (rescaled), 
i.e., right panel of Figure~\ref{fig:cospar_mock}, the constraints of the plain 2PCFs (left panel of Figure~\ref{fig:cospar_mock}) are tighter and concentrated at lower value. 
However, this does not necessarily mean that the measurement of LP from plain 2PCFs is better than the rescaled 2PCFs.

\begin{figure}
\centering
\includegraphics[width=0.49\hsize]{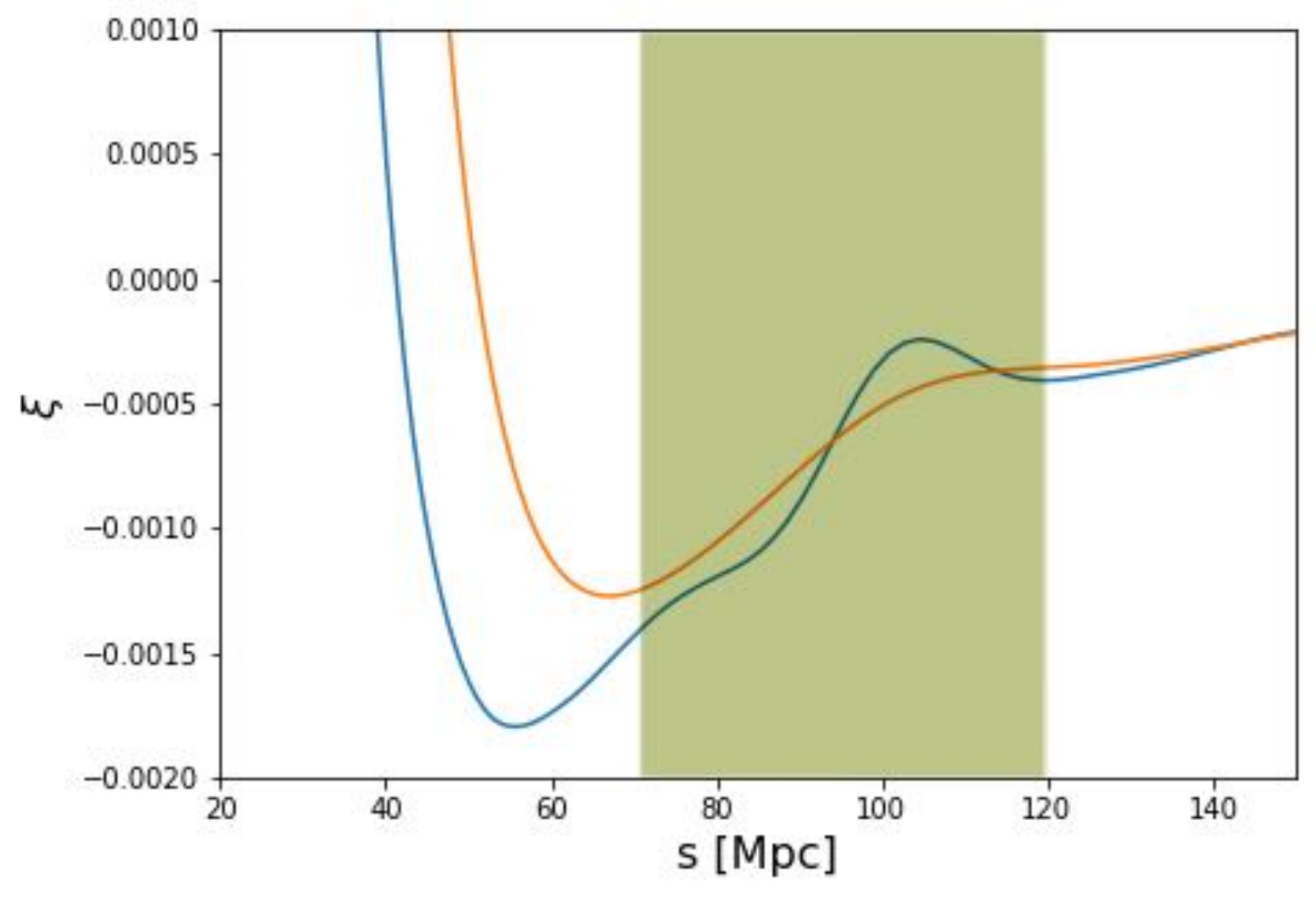}
\includegraphics[width=0.49\hsize]{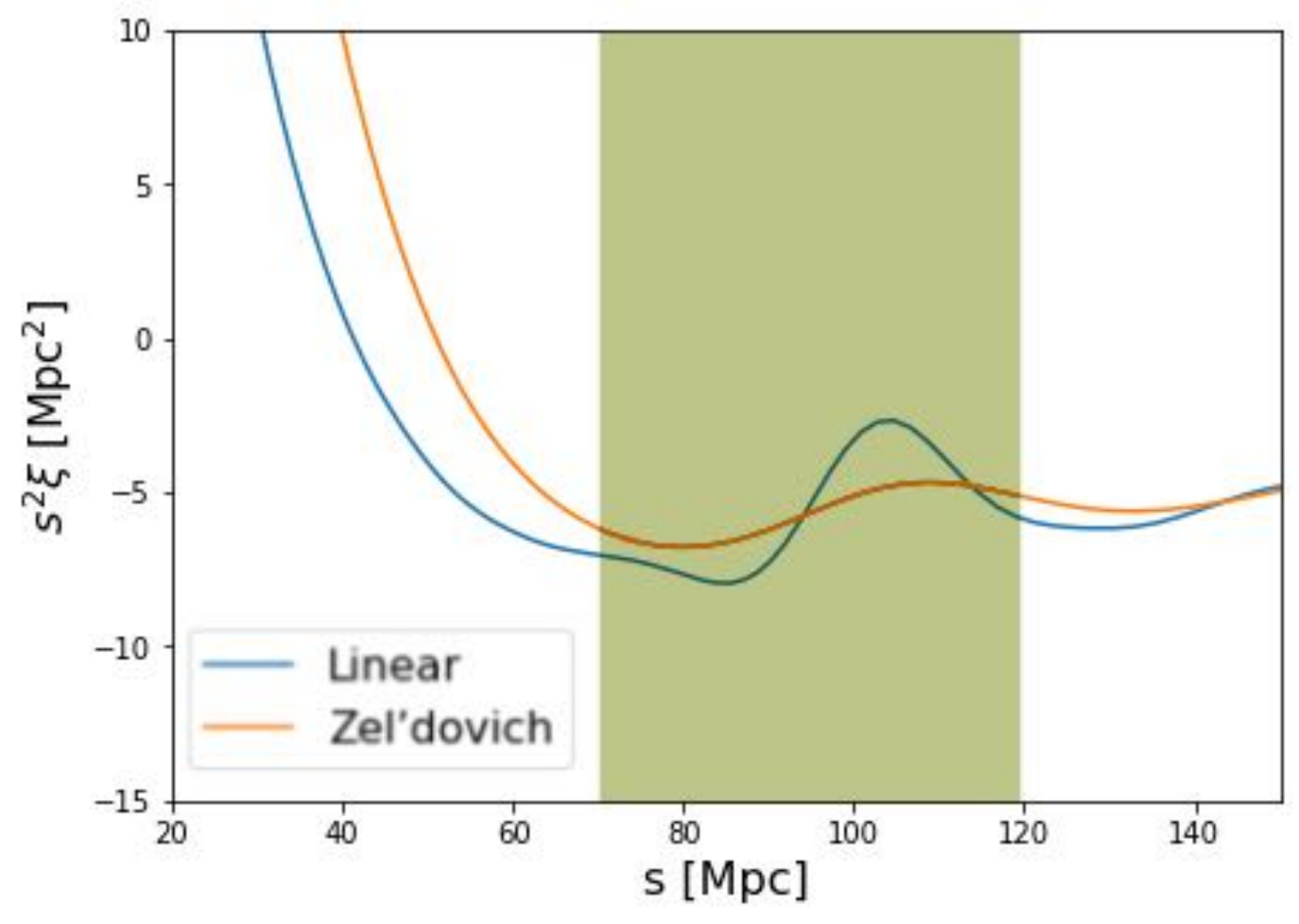}
\caption{The 2PCFs generated with extraordinary cosmology which illustrated with red star in left panel of Figure~\ref{fig:cospar_mock}. The blue lines are linear 2PCFs and the orange lines indicate the Zel’dovich approximated 2PCFs. The green shaded area refer to the range we searching for BAO peak and its associated left dip. The left panel shows plain 2PCFs ($\xi$) while the right panel shows rescaled 2PCFs ($s^2\xi$). the blue solid and orange solid line indicates linear and Zel'dovich-approximated 2PCFs, respectively. The green shadow area indicates our peak and dip search range defined as [0.7$r_{\rm d}$, 1.2$r_{\rm d}$], here $r_{\rm d}=97.3$\,Mpc.}
\label{fig:2PCFs_extraordinary}
\end{figure}

In fact, Figure~\ref{fig:2PCFs_extraordinary}, which shows the 2PCFs at cosmology shown with red star in left panel of Figure~\ref{fig:cospar_mock}, suggests that for both the linear and Zel'dovich 2PCFs, when $\Omega_{\rm m}$ and $H_0$ are larger, the BAO peak and dips are not clear or ill-behaved in the plain 2PCFs and thus non-detectable, while for the rescaled 2PCFs they are easily identified by our LP finding algorithm. 
Therefore, the exclusion of large $\Omega_{\rm m}$ and $H_0$ values by the LP measurements may be due to the weakness of the LP identification method with the plain 2PCFs.
It means that the LP constraints with the rescaled 2PCFs are more reliable. The constraints of BBN + LP (rescaled) and BBN + BAO are similar, which is consistent with the highly correlated LP and BAO measurements shown in Figure~\ref{fig:Scatter}. 
However, compared to the BBN + BAO results, the constraints from BBN + LP have a 21 per cent larger 1\,$\sigma$ confidence interval and a 0.57 $\sigma$ larger bias of the best-fitting value in $\Omega_{\rm m}$, as well as a 3.6 per cent larger 1\,$\sigma$ confidence interval and a 0.62 $\sigma$ larger bias of the best-fitting value in $H_0$. 

\begin{figure}
\centering
\includegraphics[width=1\hsize]{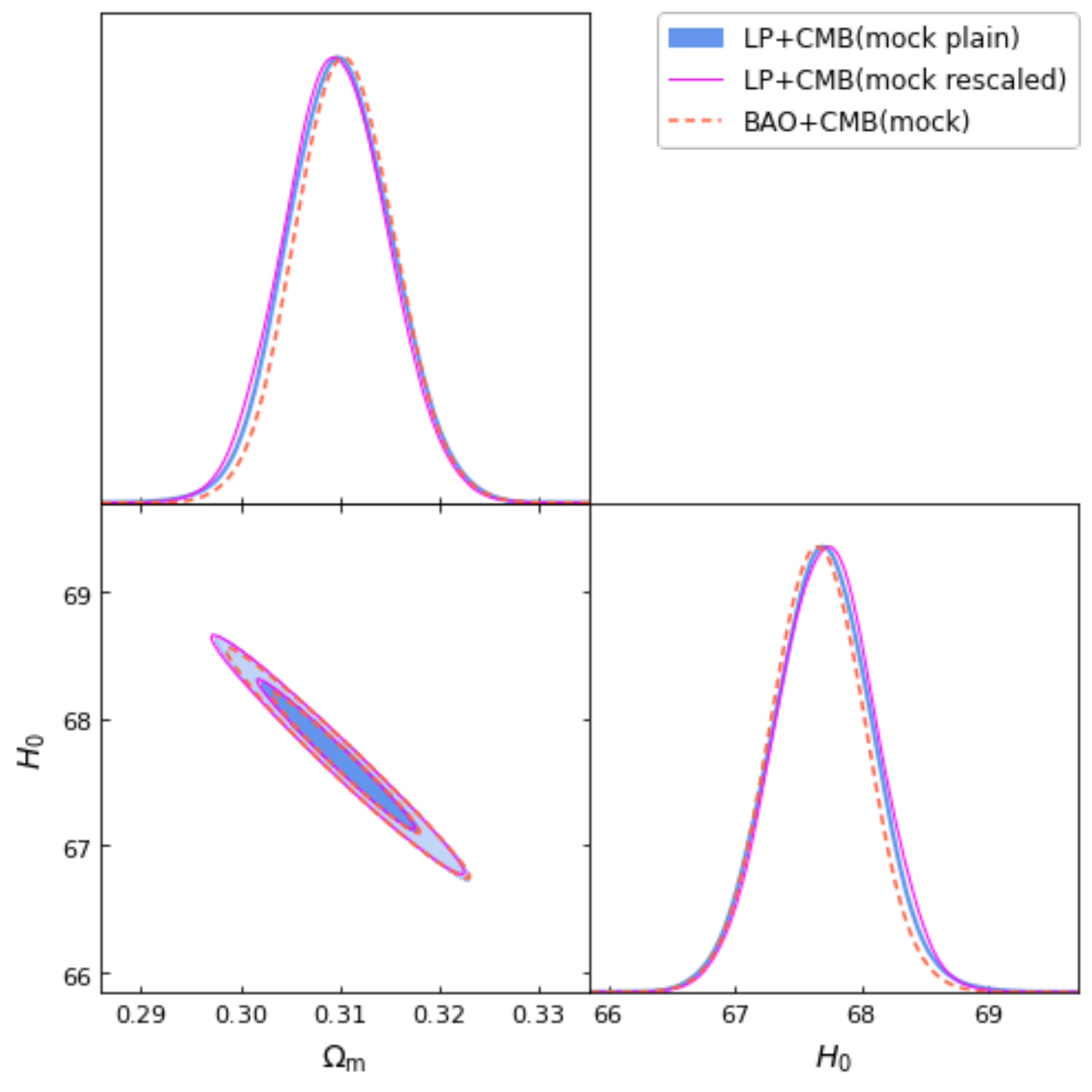}
\caption{Constraints on $H_0$ and $\Omega_{\rm m}$ based on the combination of BBN, CMB and LP (or BAO) measurements from reconstructed mocks.
The LP results from the plain and rescaled 2PCFs are shown in blue and magenta, respectively. The red contour and lines show constraints with CMB + BAO.
The blue contour and lines show constraints from CMB + LP (plain), where LP are measured from plain 2PCFs, the green contour and lines show constraints from CMB + LP (rescaled), where LP are measured from rescaled 2PCFs, while the red dashed contour and lines show that of CMB + BAO. }
\label{fig:cospar_mock_CMB}
\end{figure}

Even with the rescaled 2PCFs, the LP may still be undetectable for extreme $\Omega_{\rm m}$ and $H_0$ values. This problem can be solved by including additional observational datasets, so that the $\Omega_{\rm m}$ and $H_0$ values are better constrained. For this reason, we explore cosmological constraints with the combination of the LP measurement and Planck CMB data \citep[][]{Planck2020}. The CMB + LP and CMB + BAO results are shown in Figure~\ref{fig:cospar_mock_CMB}. 
It shows that the constraints of CMB + LP and CMB + BAO are highly consistent, even though the errors of CMB + LP are slightly larger than those of CMB + BAO. The results from the plain and rescaled 2PCFs are almost identical. 
Since the systematic error of LP measured from the plain 2PCF is smaller than that of the rescaled one (Section~\ref{subsec:systematic}), we focus on the CMB + LP (plain) results for cosmological analysis hereafter.

\section{Results}
\label{sec:results}

In this section, we present our LP measurements using the SDSS data and provide the corresponding cosmological constraints. Furthermore, we compare our results with the measurements from the template-based BAO fitting method. 

\subsection{Fitting results}

\begin{figure}
\centering
\includegraphics[width=0.98\hsize]{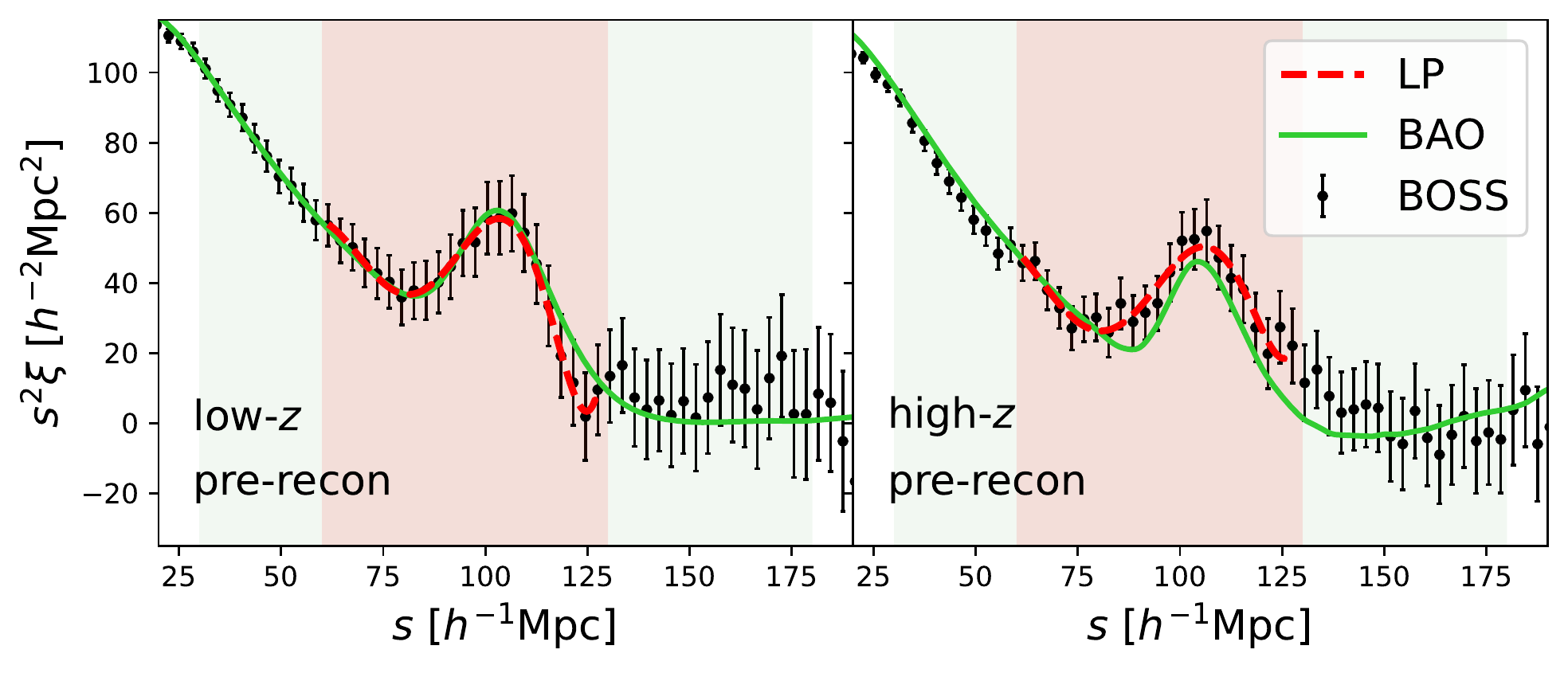}
\caption{Same as Figure~\ref{fig:2PCF_mock}, but the measured 2PCFs with error bars and the best-fitting results of the SDSS data. }
\label{fig:2PCF_SDSS}
\end{figure}

\begin{figure}
\centering
\includegraphics[width=1.\hsize]{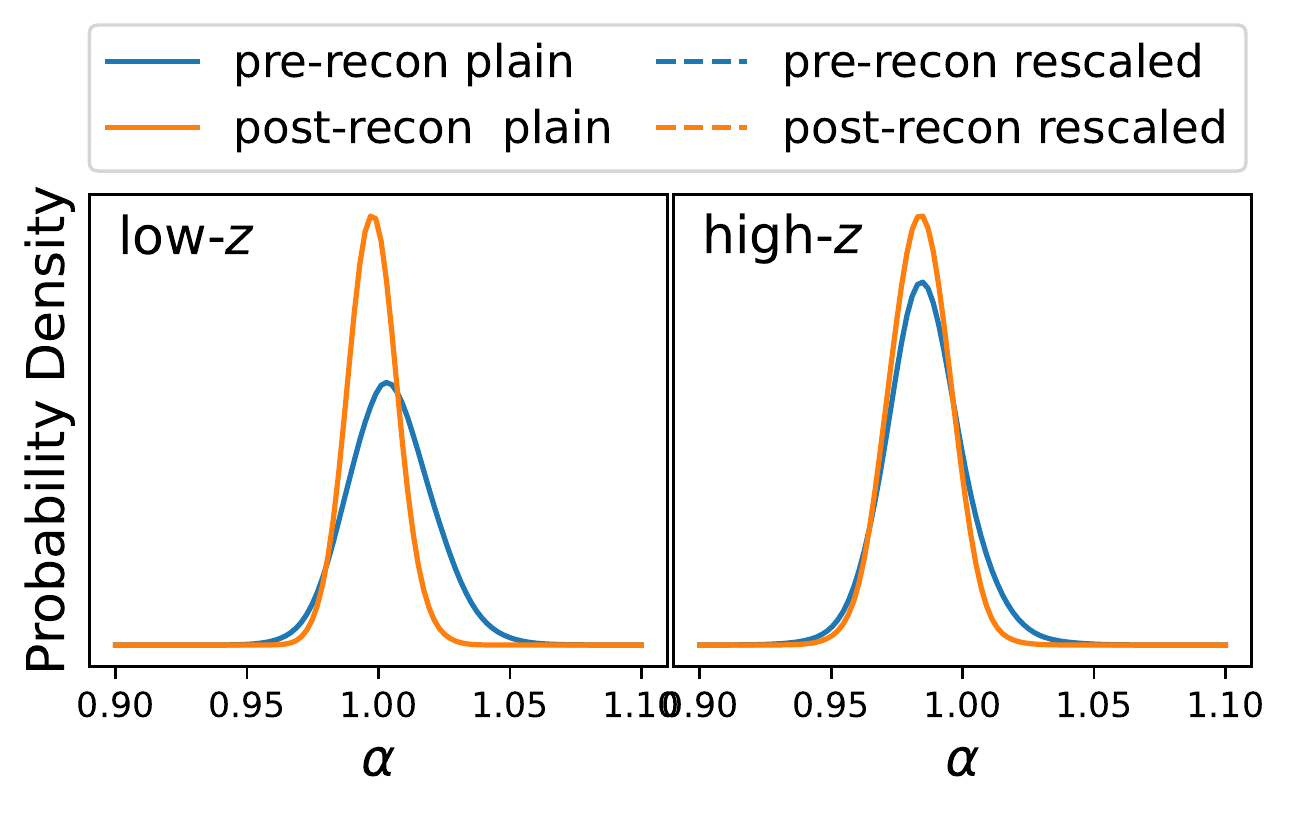}
\includegraphics[width=1.\hsize]{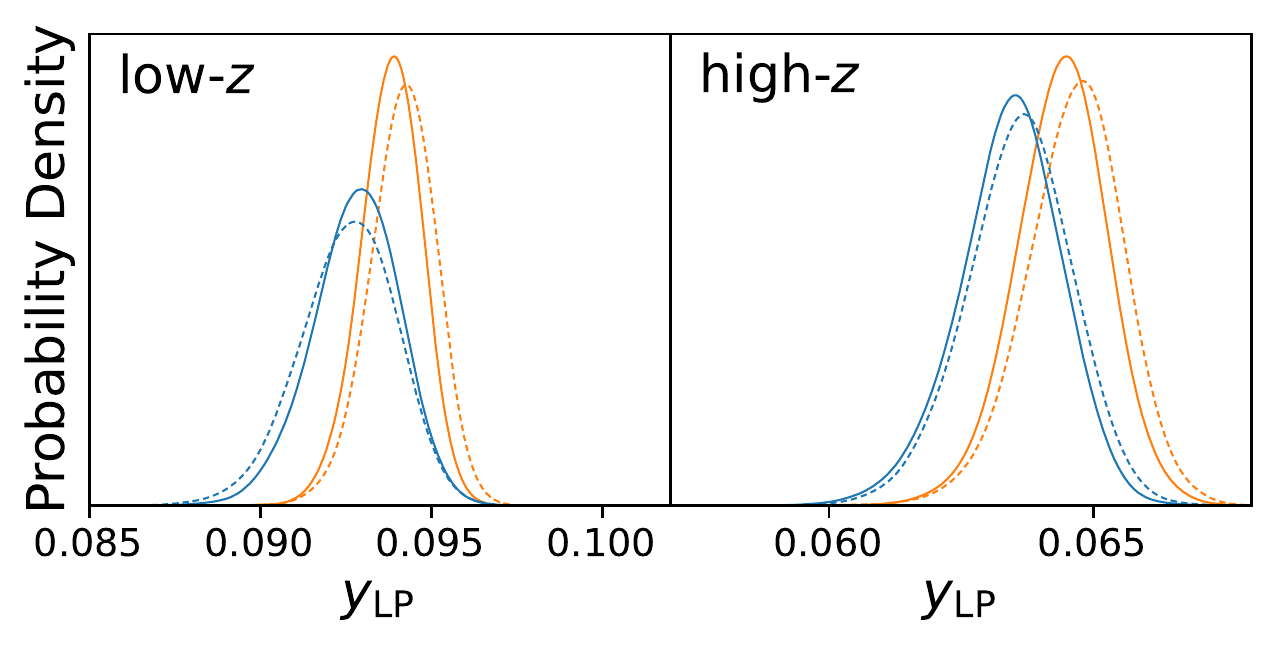}
\caption{Same as Figure~\ref{fig:posterior_mock}, but for marginalized posterior distributions of $\alpha$ and $y_{_{\rm LP}}$ fitted from SDSS data. }
\label{fig:posterior_SDSS}
\end{figure}
We perform polynomial fits for 2PCFs from the SDSS data as the fitting scheme we applied in mocks. The best-fitting 2PCFs are shown in Figure~\ref{fig:2PCF_SDSS}, along with the 2PCFs measured from the SDSS data. Both the polynomial model and the template model agree well with the observational data. The marginalized posterior distributions of $\alpha$ and $y_{_{\rm LP}}$ are illustrated in Figure~\ref{fig:posterior_SDSS}.

\begin{figure}
\centering
\includegraphics[width=0.49\hsize]{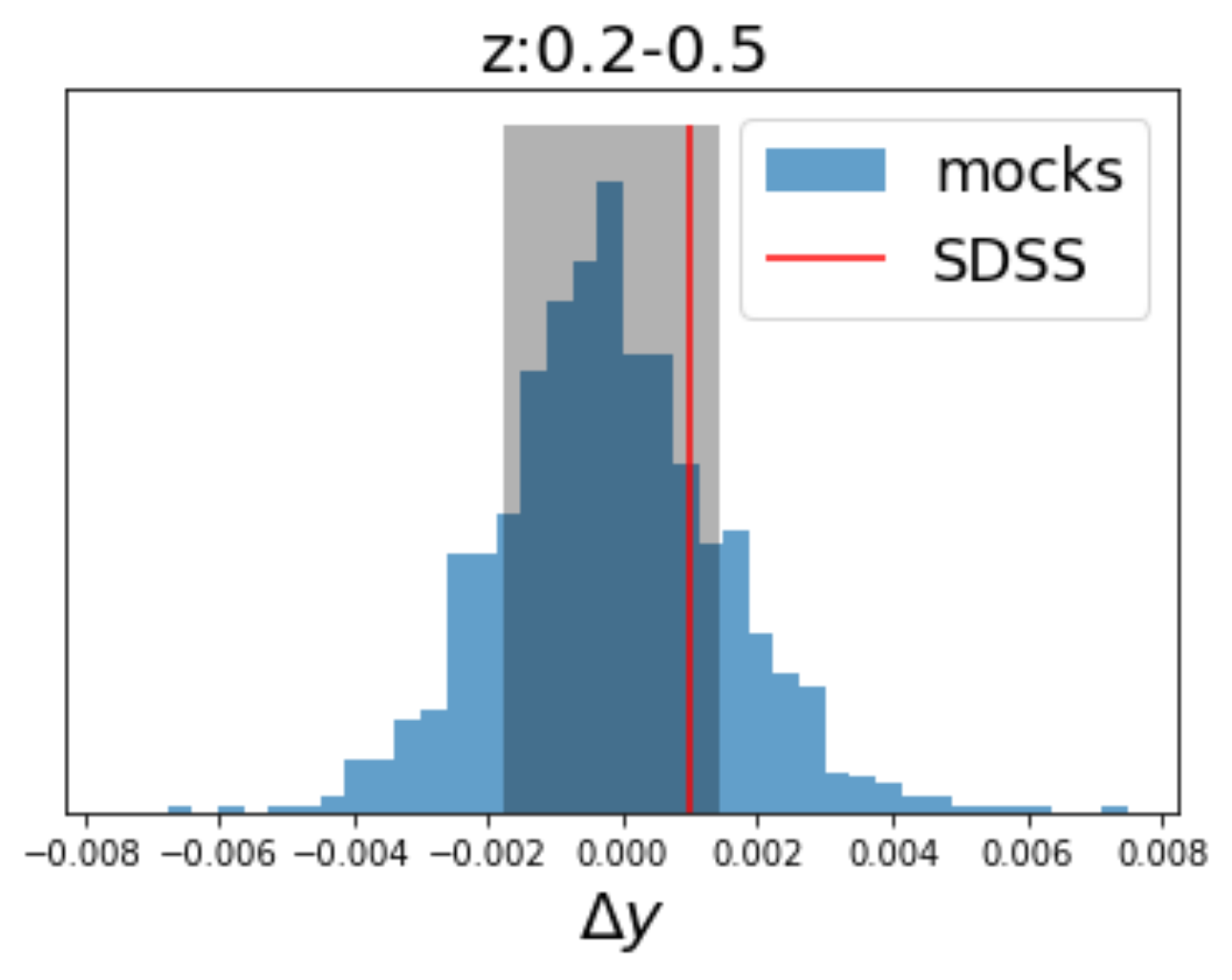}
\includegraphics[width=0.47\hsize]{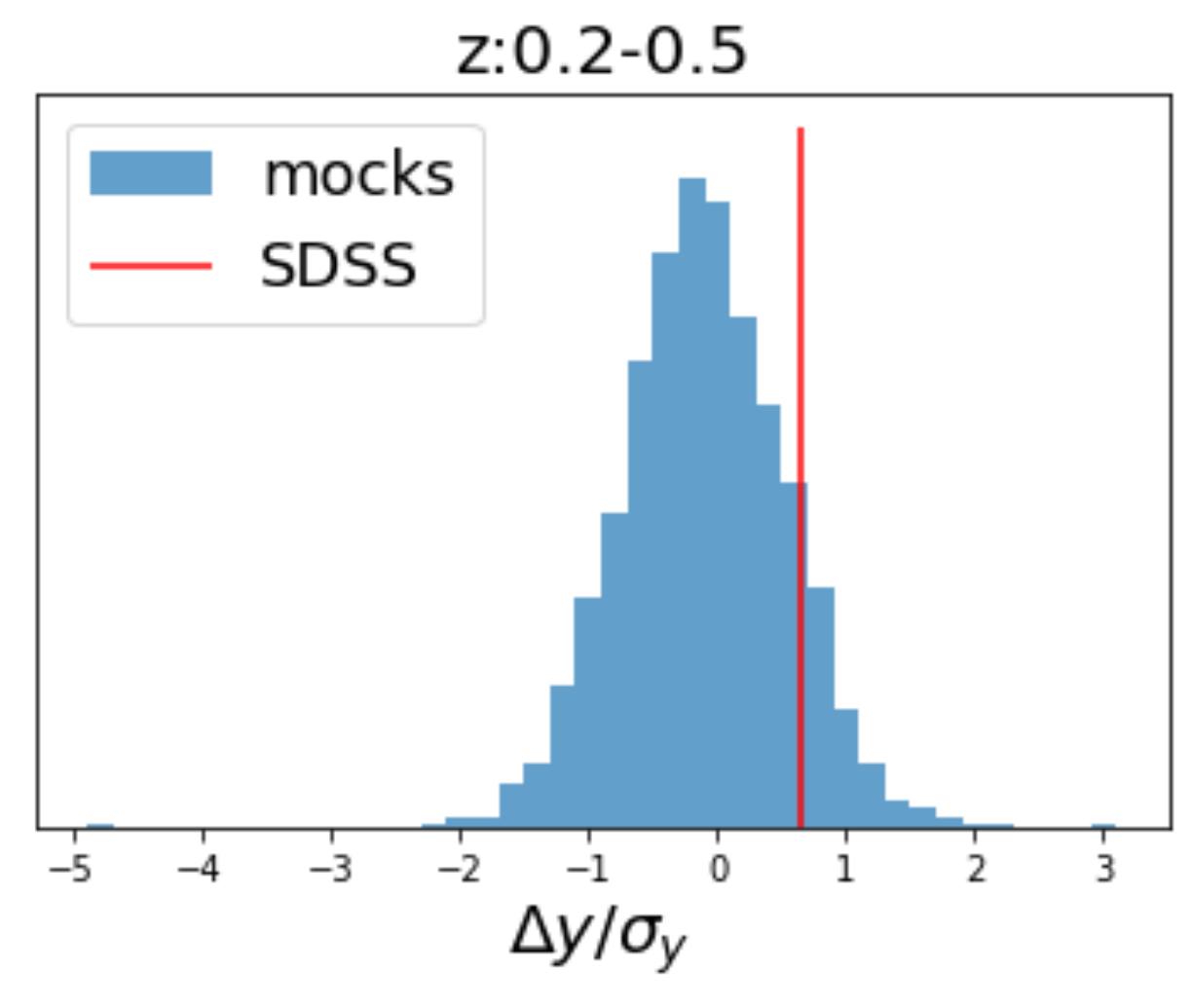}
\includegraphics[width=0.49\hsize]{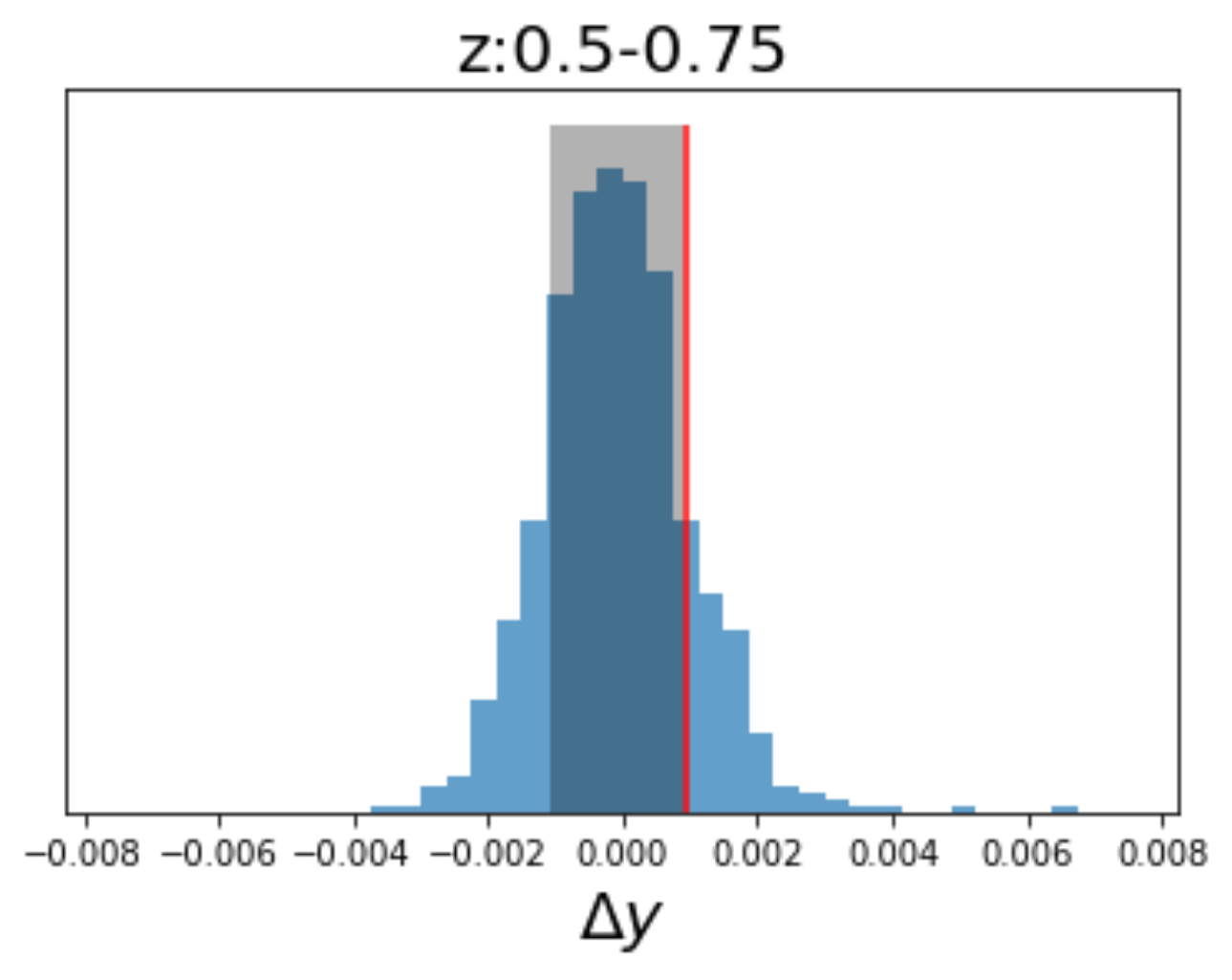}
\includegraphics[width=0.47\hsize]{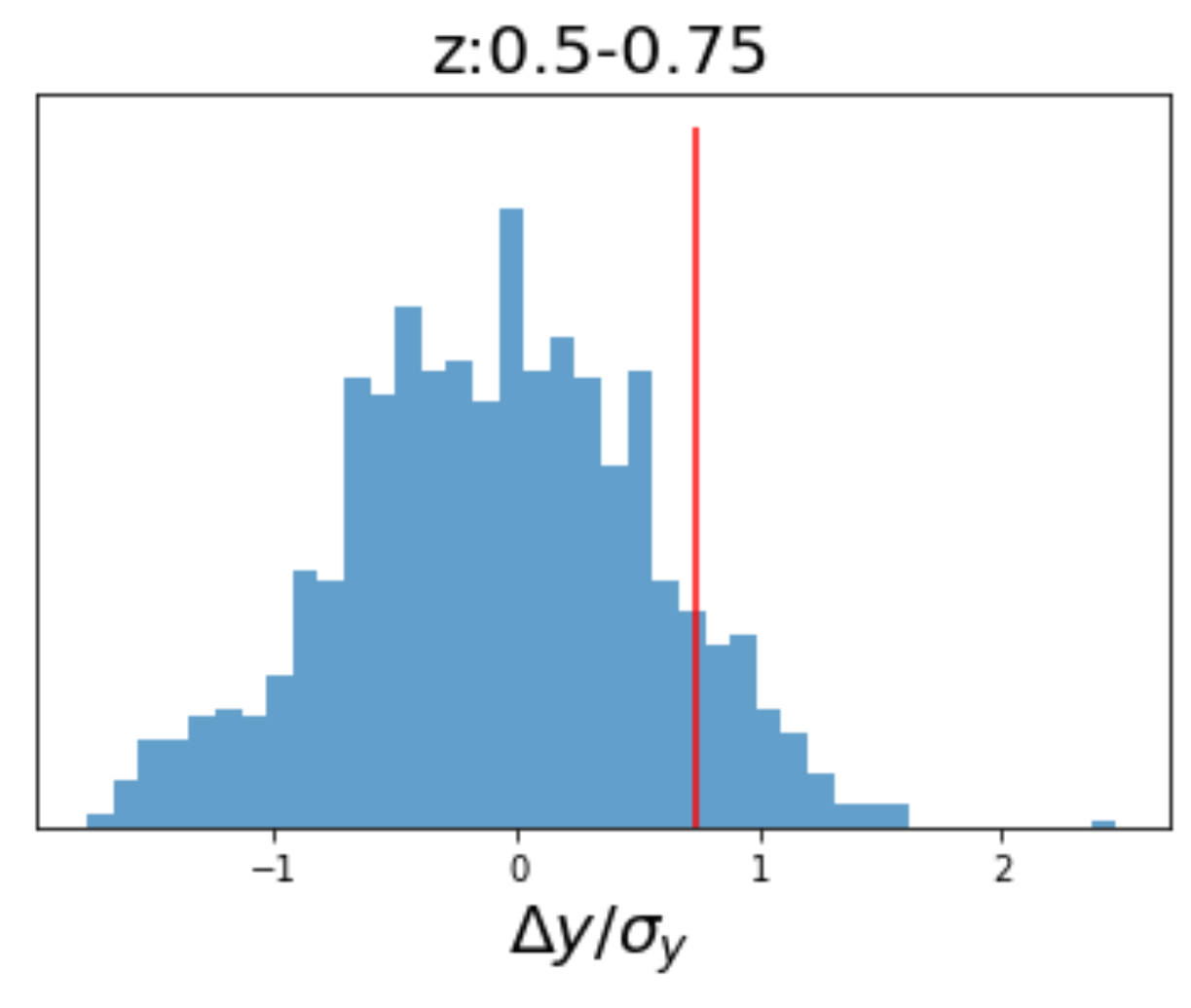}
\includegraphics[width=0.49\hsize]{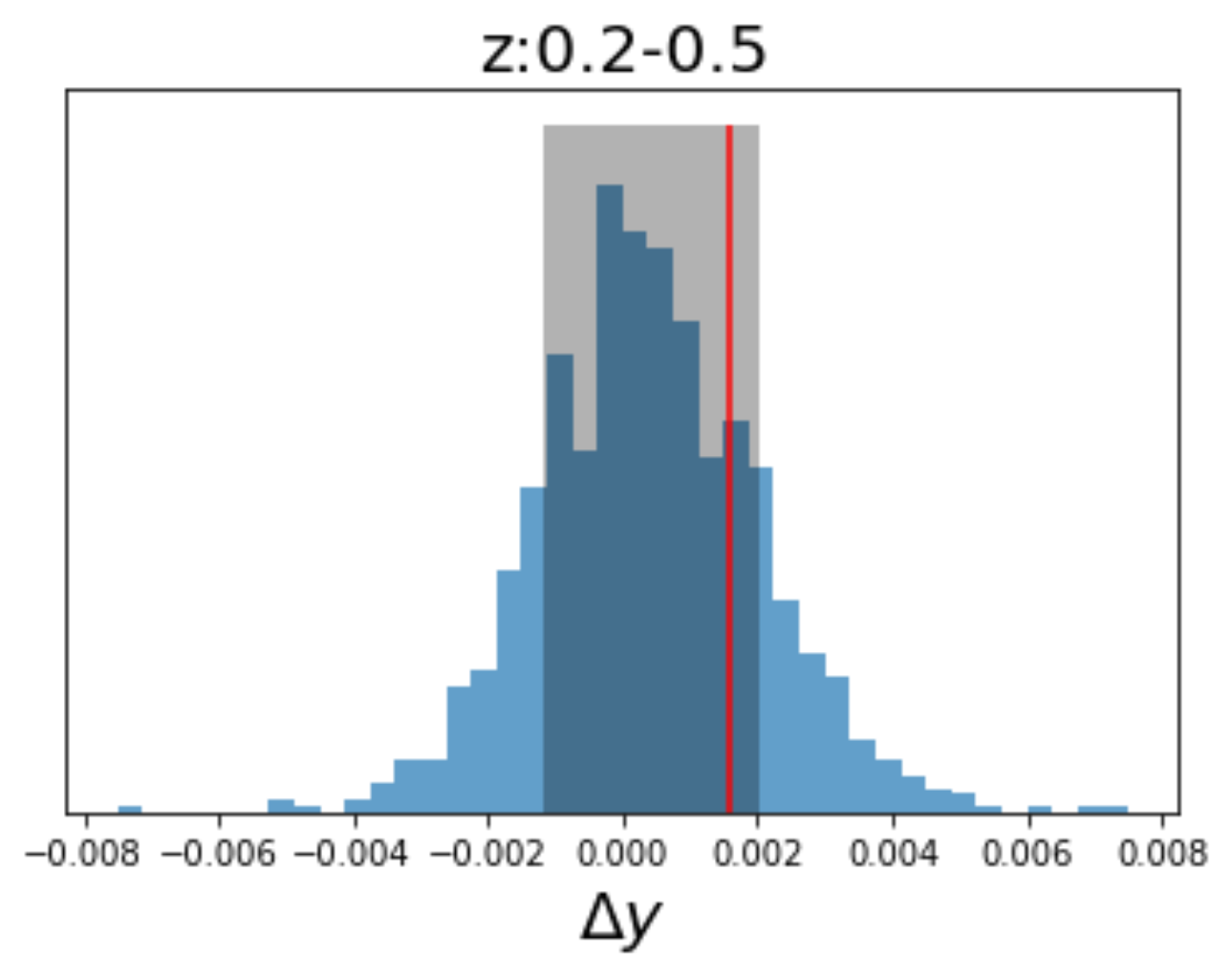}
\includegraphics[width=0.47\hsize]{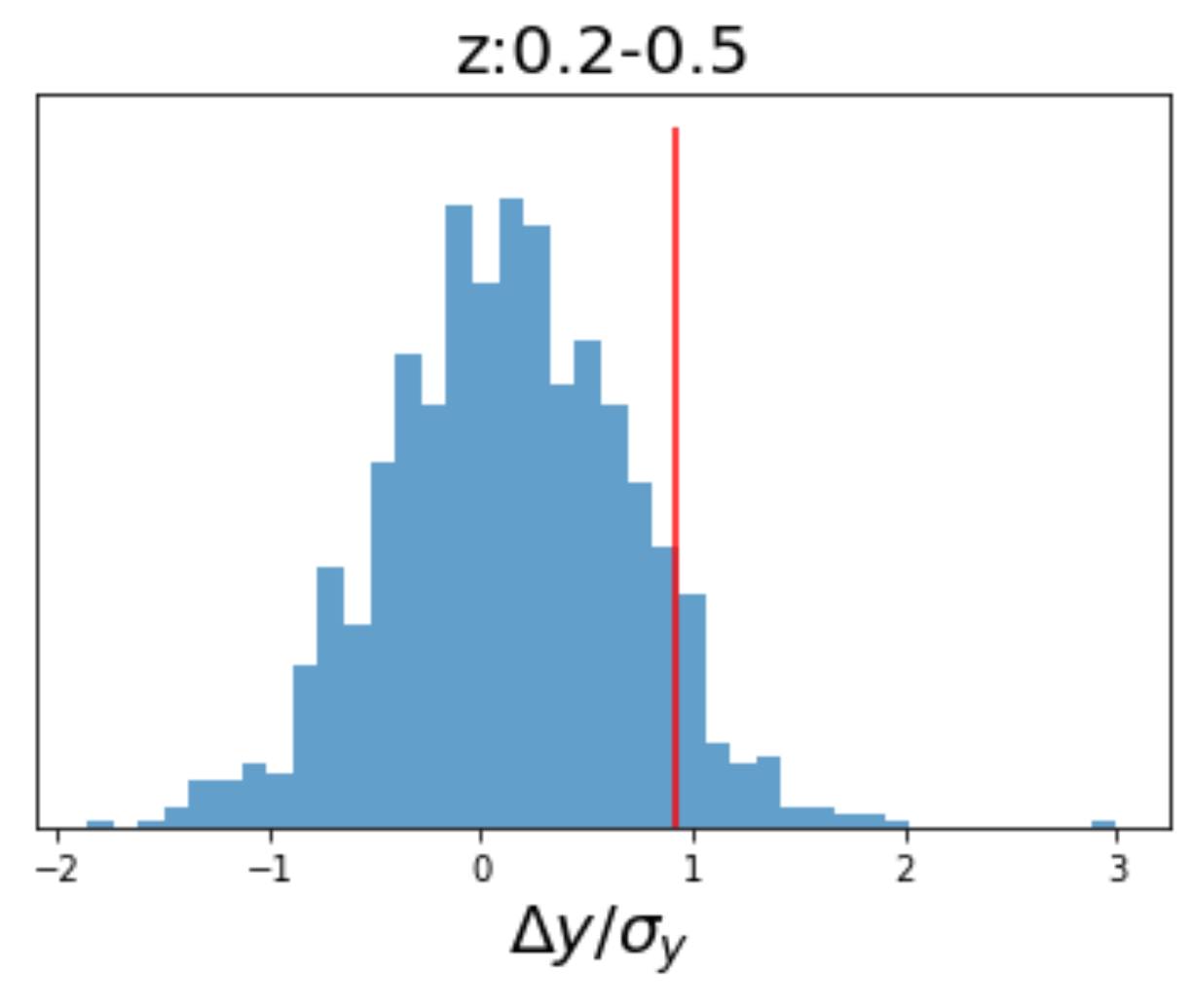}
\includegraphics[width=0.49\hsize]{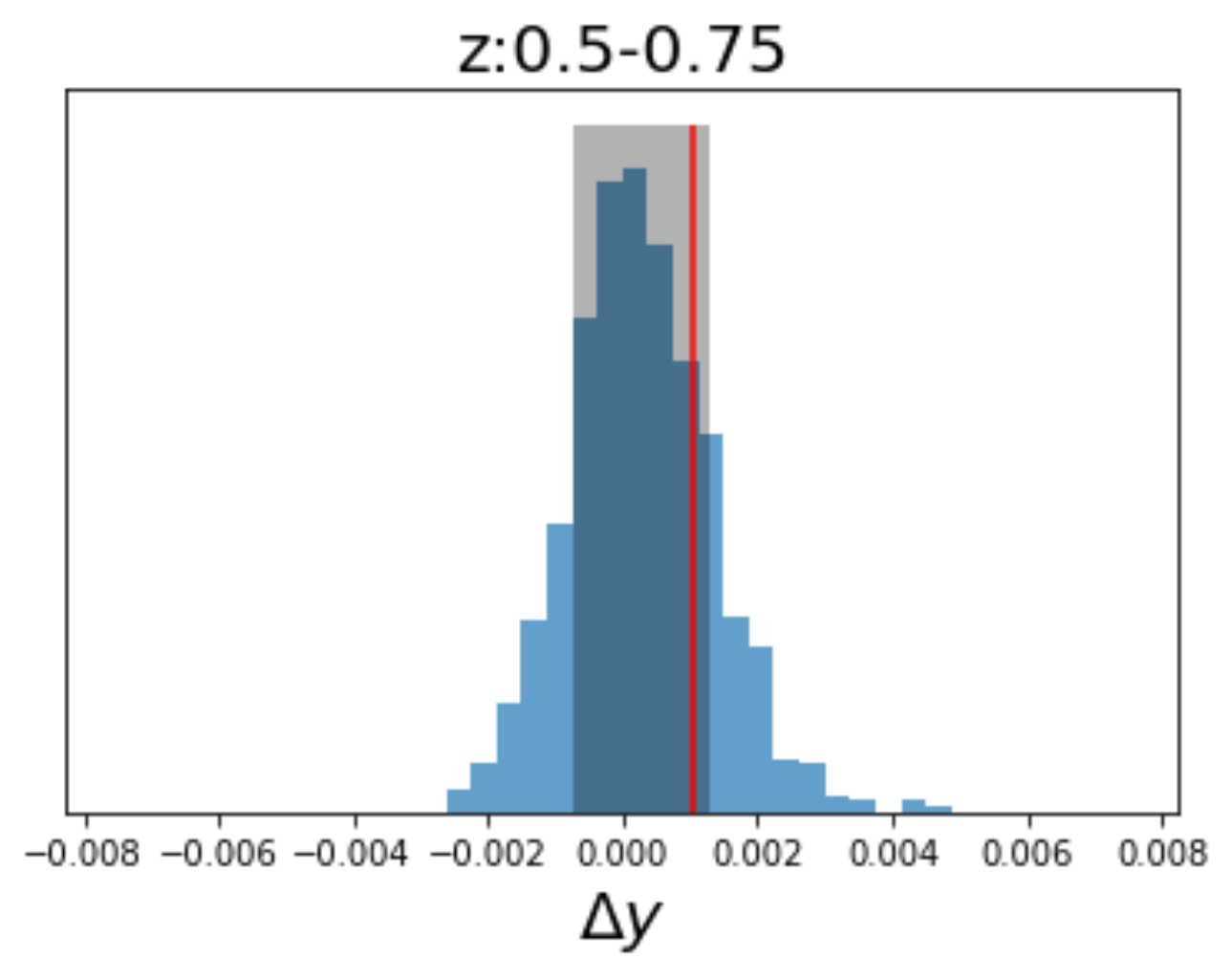}
\includegraphics[width=0.47\hsize]{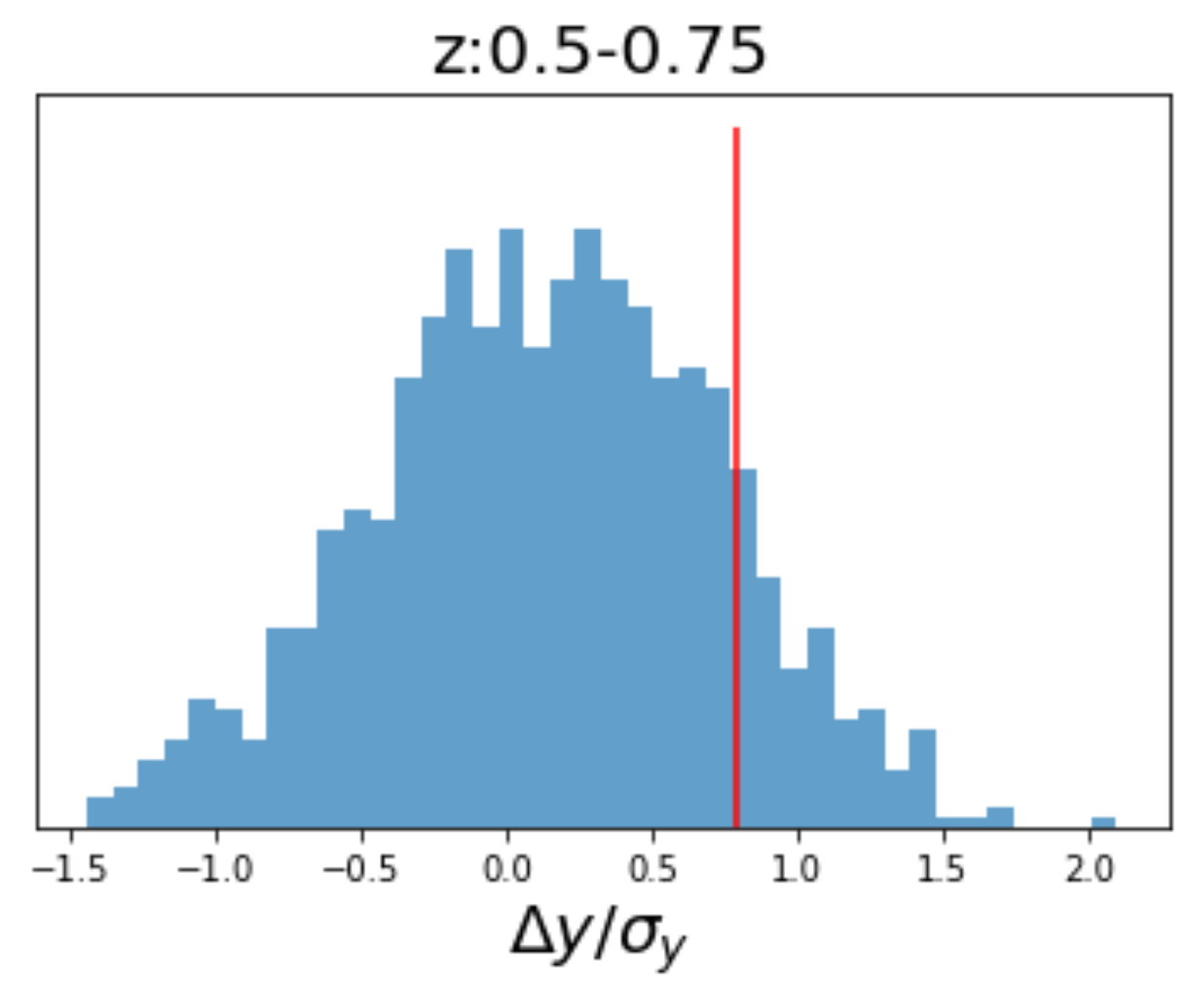}
\caption{Differences in the measured $y_{_{\rm LP}}$ ($\Delta y_{_{\rm LP}}$) and $\Delta y/\sigma_{y}$ between pre- and post-reconstruction. The blue histograms show measurements from individual mocks, the red vertical lines denote results from SDSS data and the grey shadows in left panels show the range between lower and upper 1\,$\sigma$ confidence intervals which evaluated from the 16th, and 84th percentiles of the $\Delta y$ of mocks. }
\label{fig:diff_pre_post}
\end{figure}

We find that the posterior distributions obtained from pre- and post-reconstruction data are concentrated at clearly different values, especially for posterior distributions of $y_{_{\rm LP}}$, which did not occur in posterior distributions fitted from mean 2PCFs of mocks (see Figure~\ref{fig:posterior_mock}).
This may be due to the higher level of noise in the SDSS data compared to the mock data. Figure~\ref{fig:diff_pre_post} shows the distribution of $\Delta y$ and $\Delta y/\sigma_{y}$.  We can find that the measurements of the SDSS data are all within the shaded areas and the $\Delta y/\sigma_{y}$ of the SDSS data are less than 1, indicating that the measurements from the SDSS data are not outliers compare to the corresponding mock data.

The measured $y_{_{\rm LP}}$ and $\alpha$ are presented in Table~\ref{tab:SDSS_fit_res}. We find that the statistical errors of the observed data are smaller than those of the mocks. It could be due to the nonlinear damping of the BAO peak, which has the potential to magnify the uncertainty of BAO position.
For all samples, the errors of the LP measurements are slightly higher than those of the corresponding BAO measurements.
The statistical error would be reduced by reconstruction, but the reduction in bin $z\in[0.5, 0.75]$ is not as significant as in the corresponding mocks, only 5--11 per cent, resulting in a reduction of $\sigma_{\rm comb}$ is 8--11 per cent. 
Measuring LP from rescaled 2PCFs increases the statistical error by 4--11 per cent, and increases the $\sigma_{\rm comb}$ by 3--15 per cent. 
It should be noted that additional scale-dependent observational systematic error biases the clustering on large scale significantly compare to the prediction by the mock data \citep[see][]{Ross2017}.
Although the BAO measurements are not sensitive to such observational errors, as the broad-band effects are accounted for by the nuisance parameters, they may make the LP measurements worse since $y_{_{\rm LP}}$ is directly measured from the fitted polynomial. 
This indicates that the standard BAO measurements should be more reliable than the LP measurements. 
We will examine cosmological constraints to ensure the consistency of our results in the following section.

\begin{table*}
    \centering
    \begin{tabular}{ccccccccSc}
    \hline
    SDSS DR12 & measurement & best-fitting & $\Delta$ & $\sigma$ & $\sigma_{\rm comb}$ &$\sigma_{\rm comb}$(\%) & $\frac{\chi^2}{\rm d.o.f}$ \\
        \hline
         \multirow{3}*{post  (low-z)}& $100\times y_{_{\rm LP}}$($\xi$)& 9.385 & 0.019& 0.089 & 0.091 &  0.97\%& \multirow{2}*{0.94}\\
          & $100\times y_{_{\rm LP}}$($s^2\xi$)&  9.421 & 0.044& 0.096 & 0.105 &  1.12\%\\
          \cmidrule(r){2-8}
          & $\alpha$  & 0.9977 & 0.0015& 0.009 & 0.010 &  0.95\%  & 1.06\\
         \hline
         \multirow{3}*{post  (high-z)}& $100\times y_{_{\rm LP}}$($\xi$)& 6.443 & 0.013 & 0.086 & 0.087 &  1.35\%& \multirow{2}*{1.10}\\
         & $100\times y_{_{\rm LP}}$($s^2\xi$)& 6.470 & 0.026& 0.091 & 0.095 &  1.47\% \\
         \cmidrule(r){2-8}
         & $\alpha$ & 0.9837 & 0.0036& 0.012 & 0.013 &  1.30\% & 0.65\\
         \hline
    \end{tabular}
    \caption{The measurements of $y_{_{\rm LP}}$ and $\alpha$ for observed SDSS data. }
    \label{tab:SDSS_fit_res}
\end{table*}

\subsection{Cosmological parameter measurements}

\begin{figure*}
\centering
\includegraphics[width=0.48\hsize]{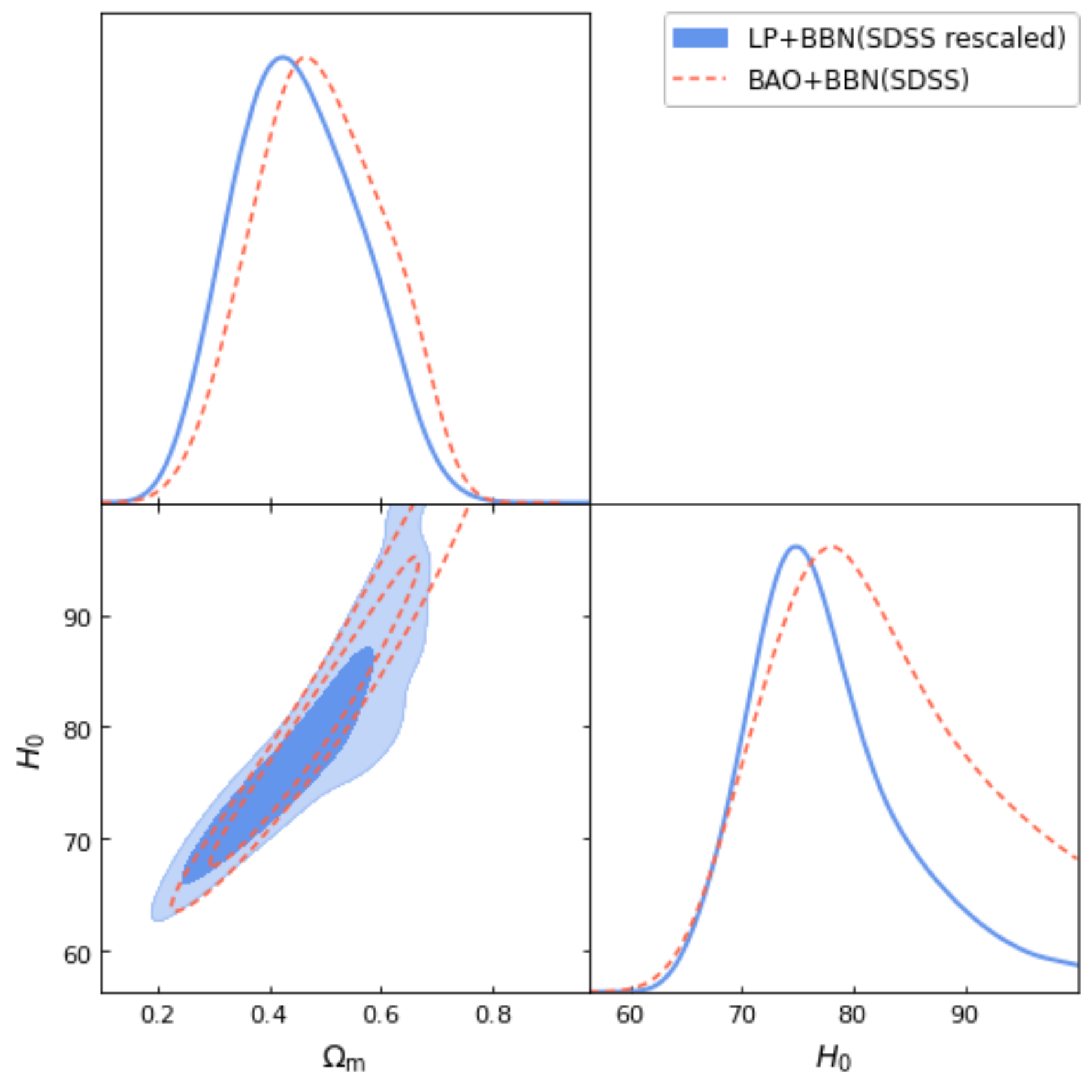}
\includegraphics[width=0.48\hsize]{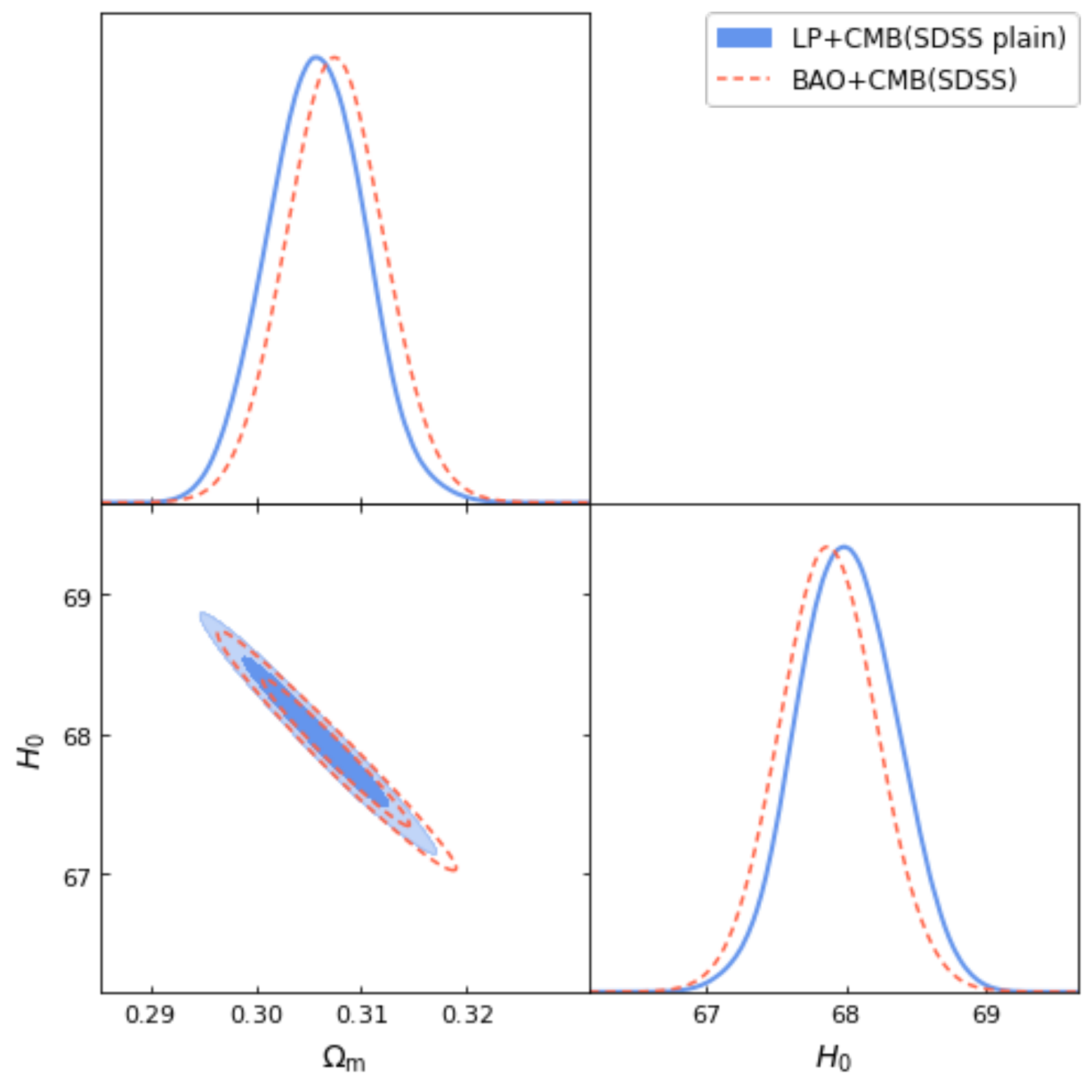}
\caption{Constraints on $H_0$ and $\Omega_{\rm m}$ based on the combinations of LP/BAO measurements from reconstructed SDSS data with BBN (left panel) and those with CMB (right panel).}
\label{fig:cospar_SDSS}
\end{figure*}

We present the constraints of the cosmological parameters in the standard flat-$\Lambda$CDM model by combining the BBN and Planck CMB data with our measurements in Figure~\ref{fig:cospar_SDSS} and Table~\ref{tab:cospar_SDSS}. 

When combined with BBN, the LP and BAO constraints are similar for $\Omega_{\rm m}$. However, there is a tension on the $H_0$ parameter. This can be explained by the fact that the LP is not be detectable with extreme $\Omega_{\rm m}$ and $H_0$ values (see Section~\ref{sec:mock_Cospar}).
The result with the LP show 3.3 per cent larger 1\,$\sigma$ confidence interval on constraints of $H_0$ than that of BAO, while the 1\,$\sigma$ constraint on $\Omega_{\rm m}$ is 0.5 per cent better.

We conclude that the LP is useful for cosmological measurements when combined with additional datasets that provide reasonably tight constraints on the parameters, such that the BAO peak is prominent.
Under this condition, the constraints from the LP are similar to those of the BAO. But the LP benefits from the fact that a cosmology-dependent template is not necessary.
We also note that with the Laguerre reconstruction method introduced by \citet[][]{Nikakhtar2021}, the determination of the LP is more robust so that the requirement on additional datasets may be released. We leave a dedicated study to the future work.

\begin{table}
    \centering
    \begin{tabular}{cScSc}
    \hline
    Data & $H_0$ (${\rm km}\,{\rm s}^{-1}\,{\rm Mpc}^{-1}$) & $\Omega_{\rm m}$ \\
    \hline
    BBN + BAO & $81.0_{-9.6}^{+7.4}$ & $0.483_{-0.112}^{+0.117}$\\
    BBN + LP (rescaled) & $77.9_{-9.3}^{+4.8}$ &$0.454_{-0.132}^{+0.116}$\\
    \hline
    CMB & 67.37$\pm 0.54$  & 0.3147$\pm 0.0074 $ \\
    CMB + BAO & $67.87_{-0.34}^{+0.35}$ &$0.3075_{-0.0046}^{+0.0046}$\\
    CMB + LP (plain) & $68.00_{-0.35}^{+0.36}$ &$0.3057_{-0.0046}^{+0.0045}$ \\
    \hline
    \end{tabular}
    \caption{The best-fitting cosmological parameters and 1\,$\sigma$ confidence intervals based on LP and BAO measured from SDSS observed data in combination with constraints from BBN and CMB.}
    \label{tab:cospar_SDSS}
\end{table}

\section{conclusions}
\label{sec:conclusions}

In this paper, we propose several novel schemes for measuring LP, such as using reconstructed data, rescaled 2PCFs ($s^2\xi$), and estimating  statistical error with Bayesian sampling.
We investigate the reliability, systematic bias, and statistical errors of these schemes with the help of approximate mock catalogues of SDSS DR12 LRG. 

Following the method described in \citet[][]{Anselmi1703}, we fit the monopole 2PCFs with a $5^{\rm th}$-order polynomial, using an $s$ bin width of $3\,h^{-1}{\rm Mpc}$ and a fitting range of 60--130\,$h^{-1}{\rm Mpc}$. The LP measurements from individuals mocks show that about 23--40 per cent of plain pre-reconstruction 2PCFs are not measurable with $S_{_{\rm LP}}$ due to the lack of a prominent BAO feature, making the LP measurements unreliable. 
This can be resolved by applying reconstruction or measuring $S_{_{\rm LP}}$ from the rescaled 2PCFs, i.e., $s^2\xi$. For instance, with reconstruction, the $S_{_{\rm LP}}$-measurable rate increases by over $\sim$95 per cent.

The $y_{_{\rm LP}}$ predicted by the Zel'dovich approximation theory, denoted as $y_{_{\rm LP, zel}}$, is more consistent with $y_{_{\rm LP}}$ measured from the mean 2PCFs of mocks for both pre- and post-reconstruction cases, compared to the theoretical value in linear theory. The systematic biases estimated using $y_{_{\rm LP, zel}}$ are less than 0.6 per cent for all measurement schemes. 
The statistical error measured with the Bayesian sampler is more stable than that of the error propagation method.
Moreover, we find the median value of the posterior distribution to be a better indicator of the best-fitting result compared to the minimum-$\chi^2$ value, by checking the consistency between measurements from individual mocks and the mean 2PCF of all mocks.  
Therefore, we use the median value and 1\,$\sigma$ confidence interval obtained from the Bayesian sampler as our final LP measurements.

Taking into account both systematic and statistical errors, $s_{_{\rm LP}}$ measured from reconstructed 2PCFs have slightly higher systematic bias compared to measurements from pre-reconstruction 2PCFs. Reconstruction can significantly reduce the statistical error by 20--30 percent, resulting in a 25--32 per cent reduction in the combined error $\sigma_{\rm comb}$. This indicates that BAO reconstruction increases the overall precision of LP measurements. 
Measuring LP from rescaled 2PCFs leads to a slight increase in both the systematic bias and statistical uncertainty of $s_{_{\rm LP}}$ for the post-reconstruction case, resulting in a 4--8 per cent increase in the combined error. 

We then compared the LP analysis with the standard BAO analysis using the mean 2PCFs of all mocks. The total relative error of $y_{_{\rm LP}}$ is approximately 20--30 per cent higher than that of $\alpha$. The cosmological results from combinations of BBN and LP measurements with the mean 2PCFs of mocks in the flat-$\Lambda$CDM cosmology indicate that, when LP is measured from plain 2PCFs, the posterior distribution of $H_0$ and $\Omega_{\rm m}$ is restricted to lower values as the BAO peak is not always prominent when scanning the cosmological parameter space.
This can be resolved by measuring the LP from rescaled 2PCFs. 
The constraints on $\Omega_{\rm m}$ and $H_0$ of both BBN + LP (rescaled) and BBN + BAO are generally consistent. However, the BBN + LP (rescaled) measurements show a 0.57 and 0.62\,$\sigma$ larger bias of the best-fitting value, as well as 21 and 3.6 per cent larger 1\,$\sigma$ confidence interval for $\Omega_{\rm m}$ and $H_0$, respectively. 

For SDSS data, the constraints from LP and BAO measurements are generally consistent. 
With the LP measurements based on quintic polynomial fitting method, the LP positions are not always detectable when cosmological parameter values are extreme, which results in unreliable constraints. 
This problem can be solved by including additional datasets such as CMB that constrain the parameter values in a relatively small volume so that the BAO peak is always prominent. In this case, the LP can produce results similar to those of traditional BAO measurements. 
When combined with CMB data \citep[][]{Planck2020}, we obtain $H_0=68.00_{-0.35}^{+0.36}$\,${\rm km}\,{\rm s}^{-1}\,{\rm Mpc}^{-1}$ and $\Omega_{\rm m} = 0.3057_{-0.0046}^{+0.0045}$ with LP measurements from plain post-reconstruction 2PCFs, and $H_0=67.87_{-0.34}^{+0.35}$\,${\rm km}\,{\rm s}^{-1}\,{\rm Mpc}^{-1}$ and $\Omega_{\rm m} = 0.3075_{-0.0046}^{+0.0046}$ with BAO measurements from the same data. 

\section*{Acknowledgements}
We acknowledge the support from the science research grants from the China Manned Space Project with No. CMS-CSST-2021-A01 and No. CMS-CSST-2021-B01. HYS acknowledges the support from NSFC of China under grant 11973070, Key Research Program of Frontier Sciences, CAS, Grant No. ZDBS-LY-7013 and Program of Shanghai Academic/Technology Research Leader.

\section*{Data availability}

All the code and data used to produce the analysis in this article are publicly available.



\bibliographystyle{mnras}
\bibliography{reference} 



\appendix
\section{The influence of fiducial cosmology}
\label{sec:sys_oms}
\begin{figure}
\centering
\includegraphics[width=1\hsize]{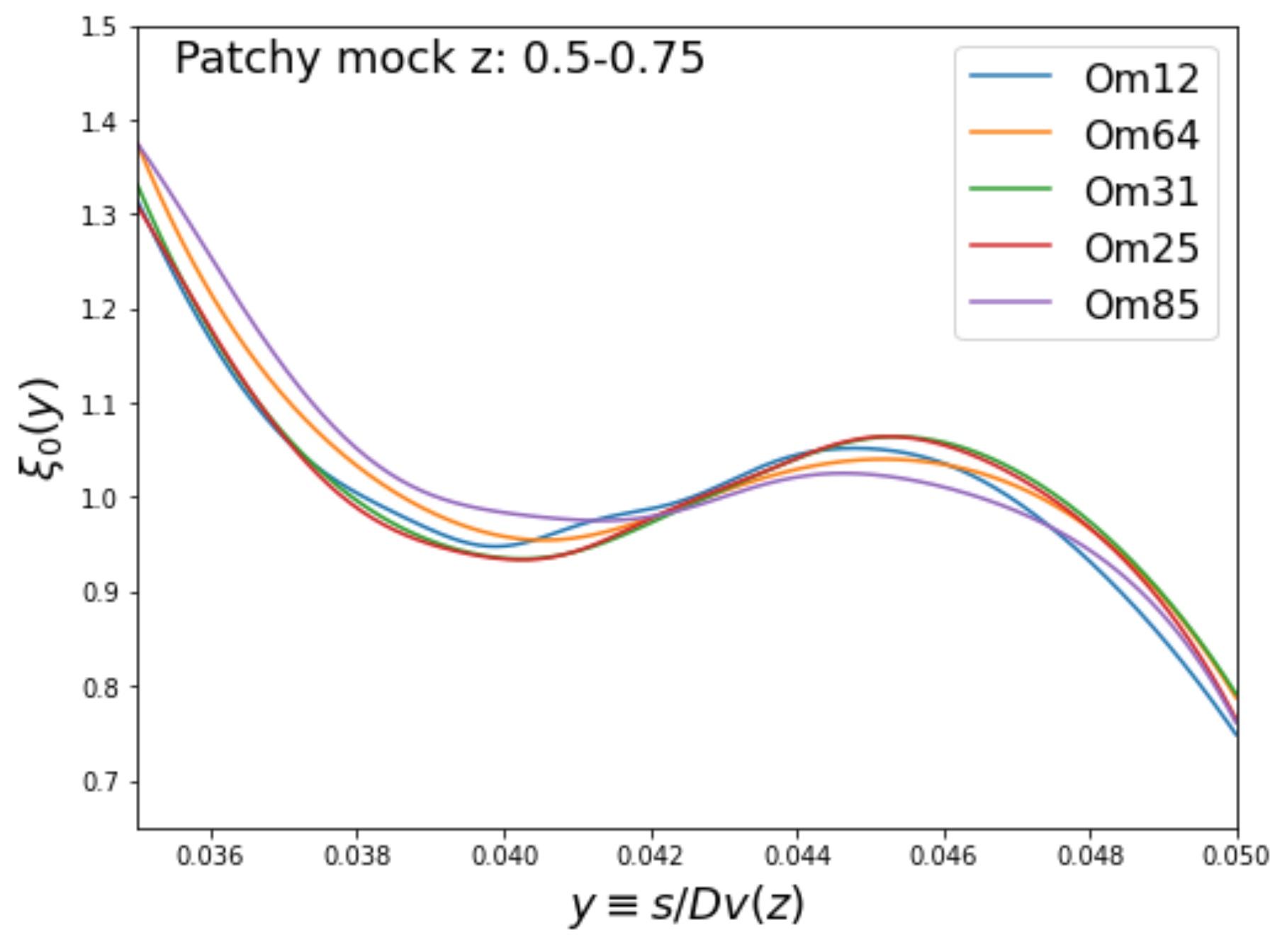}
\caption{The mean 2PCFs of Patchy mocks computed with different fiducial cosmologies.}
\label{fig:patchy_mean_Oms}
\end{figure}

We investigate the influence of fiducial cosmology, which is used to convert the measured angles and redshifts into comoving coordinates, on the LP measurement with mocks.
We measure $y_{_{\rm LP}}$ from the 2PCFs of pre-reconstruction data with different fiducial $\Omega_{\rm m}=\{0.12, 0.25, 0.31, 0.64, 0.85\}$.
$\Omega_{\rm m}$ is the key parameter used to convert the measured angular position and redshift into distance. The 2PCFs are calculated in the range of $s \in (0, 198)\,h^{-1}\,{\rm Mpc}$ with bin size of $3\,h^{-1}\,{\rm Mpc}$. 
The $y_{_{\rm LP}}$ are measured directly from 2PCFs rescaled by $D_{_{\rm V}}$. 
We show the mean $\xi(y)$ of mocks in Figure~\ref{fig:patchy_mean_Oms}. It shows that $\xi(y)$ with different fiducial cosmologies are very similar at BAO scale.

Then we run polynomial fits for these $\xi(y)$ with the fitting range of $[60 r_{\rm d}/r_{\rm d, true}$, 130 $r_{\rm d}/r_{\rm d, true}]\,h^{-1}\,{\rm Mpc}$, here $r_{\rm d}$ is the comoving sound horizon at the drag epoch calculated with corresponding fiducial cosmology.
Note that covariance matrices are rescaled by the number of mocks to ensure the statistical errors are negligible.
The $y_{_{\rm LP}}$ are estimated from searching $y_{_{\rm LP, peak}}$ and $y_{_{\rm LP, dip}}$ in the range $[ 75 r_{\rm d}/r_{\rm d, true}$, 115 $r_{\rm d}/r_{\rm d, true}]$$\,h^{-1}\,{\rm Mpc}$ with the method discussed in Section\ref{subsec:LP_measure}. 

We show the systematic bias estimated from the linear theory predicted $y_{_{\rm LP, lin}}$ and Zel'dovich approximation predicted $y_{_{\rm LP, zel}}$ in Table~\ref{tab:LP_Oms}. 
It shows that $y_{_{\rm LP}}$ is closer to $y_{_{\rm LP, zel}}$ for all cases. 
If the fiducial $\Omega_{\rm m}$ happens to be the true value, the systematic errors are less than 0.2 per cent.
Even with most extreme value of $\Omega_{\rm m}$, the systematic biases are less than 0.6 per cent. 
For the measurements from rescaled 2PCfs ($s^2\xi(y)$), the systematic errors are below 0.1 per cent if the value of $\Omega_{\rm m}$ is not far from the true one.
This indicates that the influence on  $y_{_{\rm LP}}$ from fiducial cosmology are very small, that can be negligible comparing to statistical error with the samples used in this work. 

\begin{table*}
    \centering
    \begin{tabular}{ccccccc}
        \hline
         $\Omega_{\rm m}$ & $y_{_{\rm LP}}$ ($\xi$) & $y_{_{\rm LP}}$ ($s^2\xi$)& $\Delta{y}_{lin}$($\xi$)& $\Delta{y}_{lin}$($s^2\xi$)& $\Delta{y}_{zel}$($\xi$)& $\Delta{y}_{zel}$($s^2\xi$)\\
         \hline
         0.12 & 0.06281 & 0.06269 & 1.734\% & 2.176\% & 0.387\% & 0.555\%\\
         0.25 & 0.06311 & 0.06299 & 1.257\% & 1.712\% & 0.096\% & 0.084\%\\
         0.31 & 0.06320 & 0.06308 & 1.123\% & 1.570\% & 0.231\% & 0.061\%\\
         0.64 & 0.06330 & 0.06317 & 0.961\% & 1.427\% & 0.396\% & 0.206\%\\
         0.85 & 0.06341 & 0.06319 & 0.799\% & 1.389\% & 0.560\% & 0.244\%\\
         \hline
    \end{tabular}
    \caption{The impact of $\Omega_{\rm m}$ (fiducial cosmology) in the measured $y_{_{\rm LP}}$ and its bias $\Delta y$. $\Delta{y}$ are defined as $y-y_{\rm theory}$, for $\Delta{y}_{lin}$ and the expected values of $y_{_{\rm LP}}$ are estimated from linear theory predicted 2PCFs, while $\Delta{y}_{zel}$ they are predicted by Zel'dovich-approximated 2PCFs. The columns with ($\xi$) show measured values directly from 2PCFs, while columns with ($s^2\xi$) show values from rescaled 2PCFs. Each $y_{\rm LP}$ is measured from the best-fitting mean 2PCF of the 2048 pre-reconstruction Patchy mocks. }
    \label{tab:LP_Oms}
\end{table*}


\bsp	
\label{lastpage}
\end{document}